\pdfoutput=1
\documentclass[
               superscriptaddress,
               amsmath,
               amssymb,
               reprint,
               aps]{revtex4-2}
\usepackage{graphicx}
\usepackage{dcolumn}
\usepackage{bm}
\usepackage[version=3]{mhchem}
\usepackage[caption=false]{subfig}
\usepackage{color}
\usepackage{siunitx}
\definecolor{darkblue}{rgb}{0,0,0.6}
\usepackage{hyperref}
\usepackage[utf8]{inputenc}
\usepackage[T1]{fontenc}
\usepackage{mathptmx}
\usepackage{etoolbox}
\hypersetup{colorlinks,linkcolor=darkblue,citecolor=darkblue,urlcolor=darkblue}

\makeatletter
\def\@email#1#2{%
	\endgroup
	\patchcmd{\titleblock@produce}
	{\frontmatter@RRAPformat}
	{\frontmatter@RRAPformat{\produce@RRAP{*#1\href{mailto:#2}{#2}}}\frontmatter@RRAPformat}
	{}{}
}%
\makeatother
\begin{document}

\title{Microscopic observation of two-level systems in a metallic glass model}

\author{Felix C. Mocanu}
\affiliation{Laboratoire de Physique de l'\'Ecole Normale Sup\'erieure, ENS, Universit\'e PSL, CNRS, Sorbonne Universit\'e, Universit\'e de Paris, 75005 Paris, France}
  \email{felix-cosmin.mocanu@phys.ens.fr}
\author{Ludovic Berthier}
\affiliation{Yusuf Hamied Department of Chemistry, University of Cambridge, Lensfield Road, Cambridge CB2 1EW, United Kingdom}

\affiliation{Laboratoire Charles Coulomb (L2C), Universit\'e de Montpellier, CNRS, 34095 Montpellier, France}

\author{Simone Ciarella}
\affiliation{Laboratoire de Physique de l'\'Ecole Normale Sup\'erieure, ENS, Universit\'e PSL, CNRS, Sorbonne Universit\'e, Universit\'e de Paris, 75005 Paris, France}

\author{Dmytro Khomenko}
\affiliation{Department of Chemistry, Columbia University, 3000 Broadway, New York, NY 10027, USA}

\affiliation{Dipartimento di Fisica, Sapienza Universit\`a di Roma, P.le A. Moro 2, I-00185, Rome, Italy}

\author{David R. Reichman}
\affiliation{Department of Chemistry, Columbia University, 3000 Broadway, New York, NY 10027, USA}

\author{Camille Scalliet}
\affiliation{DAMTP, Centre for Mathematical Sciences, University of Cambridge,
Wilberforce Road, Cambridge CB3 0WA, United Kingdom}

\author{Francesco Zamponi}
\affiliation{Laboratoire de Physique de l'\'Ecole Normale Sup\'erieure, ENS, Universit\'e PSL, CNRS, Sorbonne Universit\'e, Universit\'e de Paris, 75005 Paris, France}

\date{\today}

\begin{abstract}
  The low-temperature quasi-universal behavior of amorphous solids has been attributed to the existence of spatially-localized tunneling defects found in the low-energy regions of the potential energy landscape. Computational models of glasses can be studied to elucidate the microscopic nature of these defects. Recent simulation work has demonstrated the means of generating stable glassy configurations for models that mimic metallic glasses using the swap Monte Carlo algorithm. Building on these studies, we present an extensive exploration of the glassy metabasins of the potential energy landscape of a variant of the most widely used model of metallic glasses. We carefully identify tunneling defects and reveal their depletion with increased glass stability. The density of tunneling defects near the experimental glass transition temperature appears to be in good agreement with experimental measurements.
\end{abstract}

\maketitle

\section{Introduction}

The mechanical and thermal properties of glassy systems at cryogenic temperatures are determined by their low-energy excitations. In particular, at temperatures below \SI{1}{\kelvin}, where quantum effects are important, the model of tunneling two-level systems (TLS)~\cite{anderson_anomalous_1972,phillips_two-level_1987} and its extensions~\cite{ramos_low-temperature_1997} has proven to be remarkably successful in the prediction of the linear dependence in temperature $T$ of the specific heat, the $T^2$ dependence of the thermal conductivity~\cite{zeller_thermal_1971, lubchenko2003}, and the plateau in ultrasonic sound attenuation~\cite{esquinazi_tunneling_1998, buchenau_sound_2022}. Despite its success, alternative explanations for these experimental observations have been proposed~\cite{lubchenko2001,leggett_tunneling_2013}.  In particular, the interactions between TLS~\cite{burin1996}, aspects related to the scattering of phonons by TLS~\cite{zhou2015,zhou2019universal,carruzzo2019,carruzzo2021distribution}, and the collective TLS dynamics~\cite{artiaco2021} remain the subject of debate~\cite{rice_microscopic_2008}. Computational models can in principle be used to establish the relative merits and the ultimate validity of different hypotheses~\cite{heuer1993microscopic,demichelis_properties_1999,reinisch_how_2004,damart2018atomistic,khomenko_depletion_2020,kumar_density_2021}.

Beyond answering fundamental questions about the glassy state, the existence and nature of low-energy excitations in non-crystalline solids are of significant practical interest as well. In the case of superconducting circuits used for quantum computation, the tunneling TLS present in the amorphous dielectric layers of these devices are thought to be a major source of noise and decoherence~\cite{muller_towards_2019}. Similarly, the relationship of TLS defects to internal friction is an important factor for reducing the optical losses in the coatings used for glass mirrors that are part of the complex assemblies of gravitational wave detectors~\cite{birney_amorphous_2018,steinlechner_silicon-based_2018}. 

A particularly important question is what controls the density of TLS, which in turn determines the thermal and transport properties of the glass. Previous work has identified the fictive temperature ($T_f$), which is ``the temperature at which the glass sample would find itself in equilibrium if suddenly brought there from a given state''~\cite{tool} and thus efficiently encodes the thermal history of the glass, its stability and position in the potential energy landscape, as a crucial determinant of the TLS density. In fact, a reduction of roughly two orders of magnitude was observed in experiments on evaporated silicon thin films~\cite{queen2013excess}, on vapor-deposited indomethacin~\cite{perez-castaneda_suppression_2014}, and in numerical simulations of polydisperse soft-spheres~\cite{khomenko_depletion_2020}, where $T_f$ was decreased from above (hyperquenched glass) to below (ultrastable glass) the experimental glass transition temperature $T_g$.

However, the numerical results for continuously polydisperse systems~\cite{khomenko_depletion_2020} indicate a TLS density significantly larger than experiments, which suggests that other parameters might be relevant in its determination. Here, we address this question by taking advantage of a recent extension of the swap Monte Carlo algorithm~\cite{ninarello_models_2017, BerthierReichman} to computer models of metallic glasses~\cite{parmar_ultrastable_2020}, which feature both a reduced polydispersity and more realistic attractive interactions. We find the same reduction of the TLS density as a function of $T_f$ as in previous work. In addition, we find an overall depletion with respect to previous simulations of highly polydisperse soft spheres. Our work thus suggests that the use of a more realistic yet relatively simple model is sufficient to bring down the absolute density of TLS, in closer agreement with experimental measurements of typical materials~\cite{berret_how_1988, phillips_two-level_1987}. The TLS density is not a strongly universal quantity, being determined by both the microscopic parameters of the interaction potentials and the thermal history of the specific glass sample. Physically, our results confirm that increasing glass stability considerably decreases the density of TLS, thus leading to an expected decrease in dissipation in amorphous solids at low temperatures and a concomitant improvement in the material attributes associated with a variety of practical applications. 

Lastly, we revisit previous numerical protocols for TLS determination~\cite{khomenko_depletion_2020} and provide detailed insight on how the measured density of TLS depends on the exploration protocol, and on the properties of energy minima and barriers inside glass metabasins. Finally, we investigate the glass vibrational modes in order to analyze the low-frequency quasi-localized modes (QLM) found at the harmonic level. We find a weak correlation between the density of QLM and TLS, distinct from the proportionality suggested recently~\cite{ji2021toward}. Our results emphasize the diversity of low-energy excitations governing the behavior of glasses at low and cryogenic temperatures.

Our manuscript is organized as follows.
In Sec.~\ref{sec:genstrat} we present the numerical tools used to isolate and observe TLS.
In Sec.~\ref{sec:PEL} we describe the results of the potential energy landscape exploration.
In Sec.~\ref{sec:TLS} we characterise the statistical properties of the TLSs and their temperature evolution.
We discuss our results in Sec.~\ref{sec:conclusions}.

\section{Numerical methods}

\subsection{General strategy}

\label{sec:genstrat}

Our ultimate goal is to detect and characterize numerically TLS in a model metallic glass, and investigate how their number and properties evolve with glass preparation. In experiments, TLS are naturally excited at low temperature where quantum effects become relevant. However, simulating the quantum dynamics of glasses containing thousands of particles at very low temperature is prohibitively difficult, and one should find alternative approaches to identify TLS. 

Our aim is therefore to estimate $n(E)$, i.e. the number of tunneling double-well potentials (transition paths connecting distinct potential energy minima) with associated quantum splitting smaller than $E$, counted per atom and per glass sample. The TLS model~\cite{anderson_anomalous_1972,phillips_two-level_1987} postulates the small-$E$ scaling
\begin{equation}
  n(E) \simeq n_0 E + O(E^2) \ .
  \label{eq:nofE}
\end{equation}
The plateau value $n_0$ reached by $n(E)/E$ at small $E$ allows one to estimate $n_{TLS}$, the number of active TLS per atom in a typical glass sample at temperature $T_Q$ where experiments are performed, typically around \SI{1}{K},
\begin{equation}
n_{TLS} = n_0 k_B T_Q \ .
\label{eq:nTLS}
\end{equation}
The quantity $n_{TLS}$ deduced from $n(E)$ from Eqs.~\eqref{eq:nofE} and \eqref{eq:nTLS} is the key factor controlling physical properties at low temperatures.   

We provide a brief summary of our numerical strategy to measure $n(E)$ for glasses characterised by various fictive temperatures, while more detailed explanations and analysis will be presented in the following sections. There are four important steps:
\begin{enumerate}
    \item Prepare glass samples of various degrees of stability, or fictive temperatures. To do so, we generate $N_{g}$ independent equilibrium configurations in the supercooled liquid phase at different temperatures $T_f$ ranging from the mode-coupling crossover temperature $T_{\rm mct}$ down to the experimental glass transition temperature $T_g$ using the swap Monte Carlo algorithm~\cite{ninarello_models_2017}. We then form glasses by rapidly quenching the configurations to a lower ``exploration'' temperature $T_{exp}$ using conventional molecular dynamics (MD). Therefore, our glassy states at $T_{exp}$ would correspond to equilibrium supercooled liquid configurations if brought back to $T_f$, which allows us to identify $T_f$ with Tool’s fictive temperature~\cite{tool}. Varying $T_f$ is similar to varying the cooling or deposition rate in experiments. Each of the $N_g$ swap-generated equilibrium configurations defines a ``glass sample'' or a ``glass metabasin.''
    
    \item Explore using MD the potential energy landscape of each glass metabasin~\cite{Doliwa2003, Denny2003}, which contains multiple potential energy minima, or inherent structures (IS)~\cite{SW82,sciortino_potential_2005,heuer_exploring_2008}, separated by energy barriers. The low exploration temperature $T_{exp}$ is chosen to completely suppress particle diffusion over the simulated timescales, thus confining the exploration to a single glass metabasin defined by the initial configuration. At regular intervals during the dynamical trajectory simulated at $T_{exp}$, the potential energy is minimized to generate inherent structures. We call $n_{IS}$ the number of distinct IS sampled over a single glass metabasin, and $N_{IS}$ that sampled over all $N_g$ glasses.
    
    \item The pairs of inherent structures visited consecutively in the trajectory are candidate double-well (DW) potentials if the transition is observed at least once in both directions. For those we compute the minimum-energy path (MEP) connecting these pairs of IS using the nudged elastic band (NEB) method~\cite{jonsson1998nudged, henkelman_improved_2000}. We obtain a library of $N_{DW}$ double-well potentials.
    
    \item The minimum energy path provides an effective one-dimensional energy profile with two minima along which the quantum splitting, decay rate and tunneling matrix elements are estimated via a one-dimensional Schr\"odinger equation~\cite{heuer1993microscopic,khomenko_depletion_2020}. We call $N_{DW}(E)$ the total number of double wells with a quantum splitting lower than $E$, with $N_{DW}(E\to\infty)=N_{DW}$. The quantity $N_{DW}(E \sim k_B T_Q)$ is directly related to the number of two-level systems in the glass we wish to estimate.
\end{enumerate}

It is useful to decompose the number $n(E)$ of low-energy excitations per atom and per glass sample as
\begin{equation}\label{eq:nE}
    n(E) = \frac{N_{DW}(E)}{N \, N_g} = \frac{1}{N} \times
    \frac{N_{IS}}{N_{g}}\times
    \frac{N_{DW}}{N_{IS}}\times
    \frac{N_{DW}(E)}{N_{DW}} \ ,
\end{equation}
where $N$ is the number of atoms. The standard TLS model postulates that DW potentials originate from strongly localized atomic motions with a small, finite density. The number of DW in a glass is thus extensive, yielding a finite $n(E)$.
Physically, in order to be active at a given temperature $T_Q$ over an observation time $t_w$, the TLS need not only to have the relevant energy splitting $E \sim k_B T_Q$, but also to have a tunneling rate $\Gamma$ such that $\Gamma t_w \gg 1$. Because the TLS model postulates a flat distribution of $\log \Gamma$ at small $\Gamma$, the effective number of active TLS grows as $\log t_w$ at large $t_w$~\cite{anderson_anomalous_1972,phillips_two-level_1987,loponen1982}.
In summary, according to the TLS model we should expect a finite $n(E) \propto k_B T_Q \log t_w$ when $N$ and $N_g$ both diverge.

Given that our TLS detection protocol is different from the experimental one, we need to discuss how each ratio in Eq.~\eqref{eq:nE} is expected to behave when its denominator diverges, i.e. when we increase statistics. Let us first consider the ratio $N_{DW}(E)/N_{DW}$. This is the cumulative histogram of quantum splittings, which then converges to a finite limit when $N_{DW}\to\infty$. The TLS model then predicts that the remaining ratio $N_{DW}/(N_{g} \times N)$ converges to a finite value for large $N$ and $N_{IS}$. However, we discuss in Appendix~\ref{appendix:nddwnis} that this ratio, as measured in our numerical protocol, may have very different behavior depending on the nature of TLS and their interactions. In particular, because we work at fixed (and not very large) $N$, and because the $N_{g}$ glass samples are independent, the total number of sampled IS scales as $N_{IS}\propto N_{g}$ and the ratio $N_{IS}/N_g$ then converges to a finite value when $N_g\to\infty$ at finite $N$. But, as we will see, this value is so large (naturally scaling as $\exp N$) that we are not able to enumerate all the IS in a reasonable exploration time, and as a result $N_{IS}/N_g$ depends (albeit quite weakly) on exploration time in the classical MD simulation at temperature $T_{exp}$. Predicting the large scale behavior of $N_{DW}/N_{IS}$ is more intricate, see Appendix~\ref{appendix:nddwnis} for a discussion; we observe in our exploration protocol that this ratio remains of order one, and depends very weakly on the chosen exploration time.

Overall, the determination of $n_{TLS}$ weakly depends on time in both the experimental measurements where quantum dynamics is at play, and in our simulations involving classical exploration of the landscape. However, the physical origin is quite distinct in both situations, and this renders a direct quantitative comparison with experiments delicate.

In the following, we discuss in more details each step of the construction of the DW library discussing in particular in a more detailed way the behavior of the different terms in Eq.~\eqref{eq:nE}. 

\subsection{Metallic glass model}

We simulate a ternary mixture of Lennard-Jones particles (TLJ) in three dimensions. Two particles $i$ and $j$ at positions $\mathbf{r}_i$ and $\mathbf{r}_j$, separated by a distance $r_{ij} = |\mathbf{r}_i - \mathbf{r}_j|$ interact with a Lennard-Jones (LJ) pairwise potential $V(r_{ij})$, given by
\begin{equation}
    V(r_{ij}) = 4 \epsilon_{ij} \left[ \left( \dfrac{\sigma_{ij}}{r_{ij}} \right)^{12} - \left( \dfrac{\sigma_{ij}}{r_{ij}} \right)^6 \right] + S(r_{ij})  \ ,
\end{equation}
if the particle pair distance $r_{ij}/\sigma_{ij}=x_c$ is within a cutoff distance, $x_c < 2.5$, and have no interaction if their pair distance exceeds that.

To remove any resulting discontinuities from this truncation, we consider the smoothing polynomial~\cite{gutierrez_static_2015}
\begin{equation}
	S(r_{ij}) = 4 \epsilon_{ij} \left[ C_0 + C_2 \left( \frac{r_{ij}}{\sigma_{ij}} \right)^2 + C_4 \left(\frac{r_{ij}}{\sigma_{ij}} \right)^4 \right] \ .
\end{equation} 
The choice of coefficients: $C_0 = 10/x_c^6-28/x_c^{12}$, $C_2 = 48/x_c^{14}-15/x_c^{8}$ and $C_4 = 6/x_c^{10}-21/x_c^{16}$ ensures the continuity of the potential $V(r_{ij})$ and of its first two derivatives at the cutoff $x_c$. 

We consider three distinct types of particles, A (large), B (small), and C (medium), in a ratio A:B:C=4:1:1. The unit length is $\sigma_{AA}$ and the unit energy is $\epsilon_{AA}$, noted $\sigma$ and $\epsilon$ for simplicity. The interaction parameters are $\epsilon_{AB}=1.5 \epsilon$, $\epsilon_{AC}=0.9 \epsilon$, $\epsilon_{BB}=0.5 \epsilon$, $\epsilon_{BC}=0.84 \epsilon$ and $\epsilon_{CC}=0.94 \epsilon$ for the energies; and $\sigma_{AB}=0.8 \sigma$, $\sigma_{AC}=1.25 \sigma$, $\sigma_{BB}=0.88 \sigma$, $\sigma_{BC}=1.0 \sigma$ and $\sigma_{CC}=0.75 \sigma$ for the ranges. All particles have the same mass $m$. The unit time is $\tau_{LJ} = \sqrt{m \sigma^2/\epsilon}$. 

This ternary mixture is an extension of the well-known Kob-Andersen model (KA)~\cite{kob_testing_1995} with a third particle type (C) ensuring resistance against crystallization and increased efficiency for equilibration with the swap Monte Carlo (MC) algorithm, as detailed in Ref.~\cite{parmar_ultrastable_2020} where the model is referred to as KA$_2$.
We simulate $N$ particles at number density $\rho=N/L^3 = 1.35$ in a cubic box of linear size $L$ with periodic boundary conditions. This particle density is higher than the commonly investigated value $\rho = 1.2$~\cite{kob_testing_1995}. Indeed, the pairwise attraction gives rise to a liquid-gas spinodal at low enough temperature, which may intersect the glass transition line~\cite{sastry_liquid_2000}, leading to a gas-glass instability~\cite{testard_influence_2011}. The choice $\rho = 1.35$ ensures the stability of all the studied glasses, as revealed by positive values for the equilibrium pressure of the liquid down to the lowest temperatures investigated, $T=0.488$. Our model size $L\sim9.6\sigma$ is large enough to avoid finite-size effects on isolated TLS, but small enough to make it unlikely to observe more than one TLS in a single configuration.

\subsection{Supercooled liquid dynamics and glass preparation}

Once the details of the model have been established, one has to estimate the relevant temperatures that govern its glassy dynamics. These are the onset temperature $T_o$, the mode-coupling temperature $T_{\rm mct}$, and the laboratory glass transition temperature $T_g$.

\begin{figure}[t]
	\includegraphics[width=\linewidth]{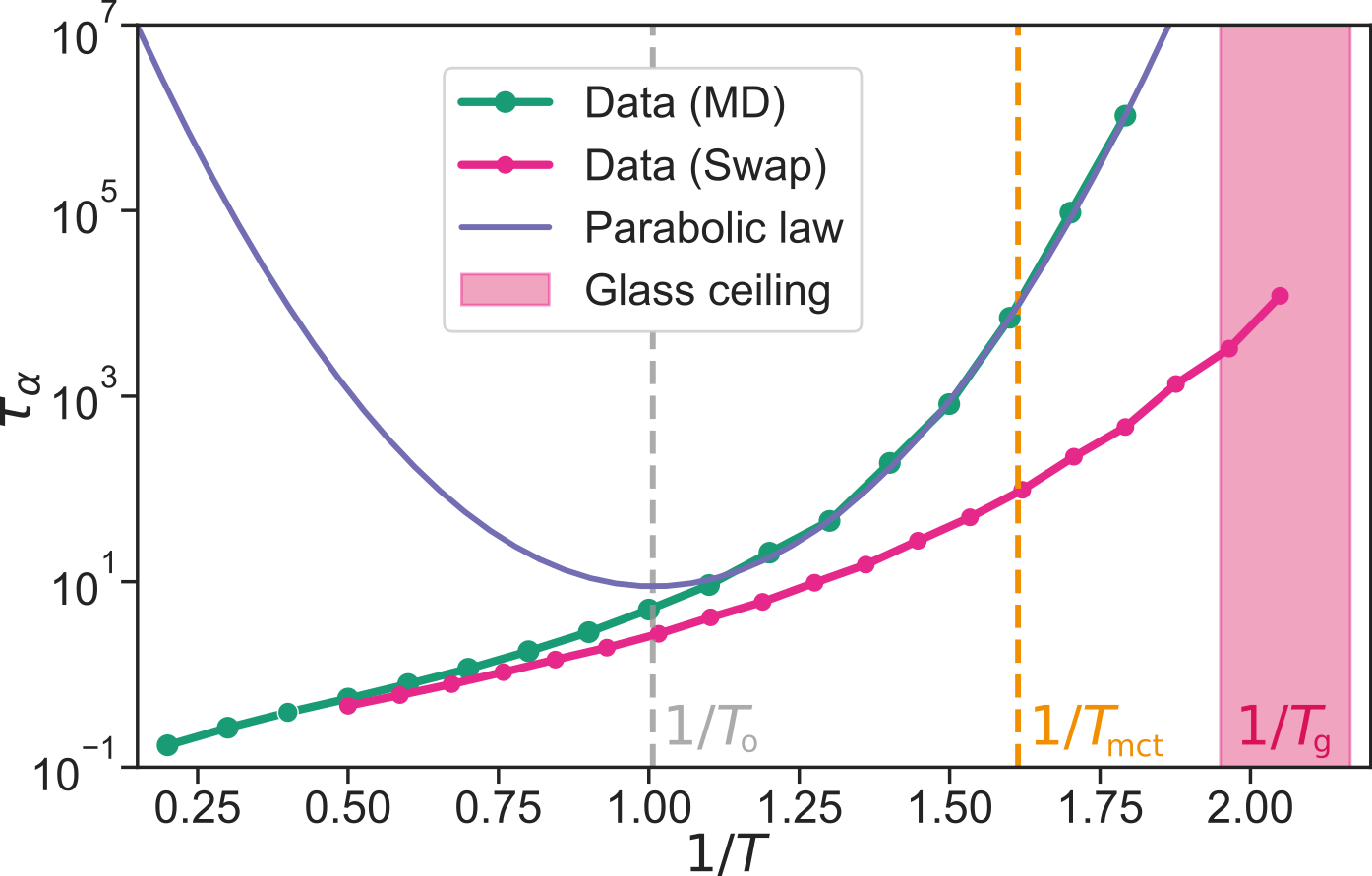}
	\caption{Angell plot of relaxation time $\tau_{\alpha}$ versus inverse temperature measured in physical molecular dynamics (MD, green) and equilibration particle-swap dynamics (Swap, magenta). The onset $T_o \simeq 1.0$ (grey), and mode-coupling crossover $T_{\rm mct}\simeq 0.62$ (orange) temperatures are located with dashed lines. The parabolic fit (purple) is used to extrapolate the relaxation time data at lower temperature. The shaded region between a VFT and parabolic law extrapolation for $T_g$ locates the ``glass ceiling''.}
	\label{fig:angell}
      \end{figure}
      
      To this aim, standard molecular dynamics has been carried out using the LAMMPS code~\cite{plimpton_fast_1995}. A time-step of $dt=5\times10^{-3} \tau_{LJ}$ together with a Nos\'e-Hoover thermostat was used to sample the canonical (NVT) ensemble.

A hybrid swap MC/MD dynamics was also carried out using a modified version of LAMMPS, as detailed in~\cite{berthier_efficient_2019}. Based on previous work~\cite{parmar_ultrastable_2020}, we have found that short blocks of 10 MD steps with $dt=5\times10^{-3} \tau_{LJ}$ interspersed with blocks of 2100 ($1.75 \times N$) attempted particle swaps produce the optimal speed-up of the dynamics of this system in the deeply supercooled regime. We recall that both the standard and swap dynamics sample the same equilibrium thermodynamics of the system, provided proper decorrelation can be achieved within the simulation time window~\cite{ninarello_models_2017}.

In order to monitor structural relaxation during standard and swap MC dynamics, we have estimated the structural relaxation time $\tau_{\alpha}$ from the self-intermediate scattering function $F_{\rm s}(q=7.2, t=\tau_{\alpha}) = 1/e$. The relaxation time $\tau_{\alpha}$  is plotted against inverse temperature in the manner of an Angell plot in Fig.~\ref{fig:angell}.
      
The onset of glassy dynamics is located at $T_o \sim 1.0$, where the relaxation time data of MD dynamics starts to deviate from a log-linear Arrhenius dependence valid at higher temperatures. To obtain $T_{\rm mct}$, we fit the relaxation times to a power-law behavior, $\tau_\alpha \propto (T-T_{\rm mct})^{-\gamma}$, in a region of moderate supercooling, which results in an empirical estimate of $T_{\rm mct}\sim0.62$. Finally, the laboratory glass transition temperature $T_g$ is the temperature at which $\tau_{\alpha}(T_g) = 10^{12}\tau_o$, in microscopic units. Simulations of such long time scales are not tractable, so this temperature is obtained from extrapolations of the available data. Two extrapolations have been used: the Vogel-Fulcher-Tammann law~\cite{ediger96} and the parabolic law~\cite{Elmatad}. These extrapolations define a range for the laboratory glass transition temperature of $T_g\sim0.46 - 0.51$ which we refer to as a ``glass ceiling'', because experiments cannot easily access temperatures lower than $T_g$. Swap dynamics accelerates the relaxation by up to 8 orders of magnitude at $T_g$, and enables the preparation of deeply supercooled configurations of the TLJ model, equilibrated in the liquid state near the glass ceiling. This allows us to explore for the present model of a metallic glass a broad range of fictive temperatures.       

\begin{table*}
	\begin{ruledtabular}
	
		\begin{tabular}{  c|c|c|c|c|c|c|c }
			
			&       \multicolumn{3}{c|}{Dataset 1}                                                      &               \multicolumn{4}{c}{Dataset 2}                                            \\
			$T_f$                    & $0.617$             & $0.558$              & $0.509$             & $0.617$              & $0.558$             & $0.509$             & $0.488$             \\
			$N_{g}$                  & $16$                & $64$                 & $256$               & $128$                & $256$               & $512$               & $1024$              \\
			$\tau_{cg}$              & $11$                & $17$                 & $43$                & $2$                  & $3$                 & $18$                & $31$                \\
			$N_{cg}$                 & $160000$            & $40000$              & $10000$             & $100000$             & $100000$            & $100000$            & $100000$            \\
			$\tau_{total}$           & $1760000$           & $680000$             & $430000$            & $200000$             & $300000$            & $1800000$           & $3100000$           \\
			$N_{IS}$                 & $36247$             & $23287$              & $30180$             & $70326$              & $88112$             & $119159$            & $117780$            \\
			$N_{DW}$                 & $10039$             & $8359$               & $12117$             & $11841$              & $20692$             & $41342$             & $38027$             \\	
			$N_{TLS}$                & $40$                & $20$                 & $41$                & $34$                 & $61$                & $137$               & $125$               \\
			$N_{IS}/N_{g}$           & $2.27\times10^{3}$  & $3.64\times10^{2}$   & $1.18\times10^{2}$  & $5.49\times10^{2}$   & $3.44\times10^{2}$  & $2.33\times10^{2}$  & $1.15\times10^{2}$  \\	
			$N_{DW}/N_{IS}$          & $2.77\times10^{-1}$ & $3.59\times10^{-1}$  & $4.01\times10^{-1}$ & $1.48\times10^{-1}$  & $2.35\times10^{-1}$ & $3.47\times10^{-1}$ & $3.23\times10^{-1}$ \\
			$N_{TLS}/N_{DW}$         & $3.98\times10^{-3}$ & $2.39\times10^{-3}$  & $3.38\times10^{-3}$ & $2.87\times10^{-3}$  & $2.96\times10^{-3}$ & $3.31\times10^{-3}$ & $3.29\times10^{-3}$ \\
		\end{tabular}
	\end{ruledtabular}
	\caption{Summary of the parameters used and the obtained statistics in the landscape exploration of the metallic glass model. Two independent data sets are used. The ratios are evaluated at the end of the exploration protocol $t = \tau_{total}$.}
	\label{table:exploration_results}
\end{table*}

\subsection{Landscape exploration via classical molecular dynamics}      

To explore the energy landscape we used an in-house standard MD code in order to have better control over the exploration workflow. The equations of motion were integrated with a smaller timestep of $dt=2.5\times10^{-3} \tau_{LJ}$ to ensure a good quality of the structures sampled and to adequately distinguish between distinct IS. Initial velocities are chosen from the Maxwell-Boltzmann distribution at $T_{exp}=0.4 < T_g$, and temperature is then kept constant by a Berendsen thermostat~\cite{frenkelsmit}. This very low exploration temperature is meant to confine the exploration to the glass metabasin selected by the initial conditions, while still allowing the system to cross enough energy barriers within the metabasin. We have carefully checked that, because standard MD is essentially arrested at those temperatures (Fig.~\ref{fig:angell}), no diffusion is observed. The mean-squared displacement reaches a plateau over a microscopic time and does not grow above it. Hence, the glass metabasin exploration is fully driven by the thermal vibrations of the particles at $T_{exp}$, and no significant change in the solid structure is detected.
      
Once every $\tau_{cg}/dt$ MD steps, we use the current configuration as initial state to minimize the energy via a conjugate gradient algorithm, to obtain an inherent structure. We repeat this procedure $N_{cg}$ times until a large library of IS is obtained, reaching a simulation time $\tau_{total} = N_{cg}\tau_{cg}$.

Once a library of IS has been constructed, we need to select IS pairs as likely candidates to be connected by a path having a double-well potential shape. To this aim, we select pairs of IS that appear consecutively in the exploration dynamics, and record the number of transitions between IS $\alpha$ and IS $\beta$ in a matrix $T_{\alpha\to\beta}$, where the labeling is ordered by energy (see Fig.~\ref{fig:matrices}). If a transition occurs between $\alpha$ and $\beta$ in both directions at least once ($T_{\alpha\to\beta} \geq 1$ and $T_{\beta\to\alpha} \geq 1$), we consider that the pair $(\alpha,\beta)$ is a good candidate double-well potential and we attempt to evaluate the MEP between them. 
      
Once a library of candidate transitions is obtained from the minima, the MEP is found using the string method of Ref.~\cite{e_string_2002}, which is implemented in LAMMPS as a modification of the nudged-elastic band (NEB) approach~\cite{henkelman_improved_2000}. We used $N_{images}=64$ images of the system to interpolate between the two IS. The images are connected by harmonic springs with a spring constant $\kappa=0.125$ $\epsilon \sigma^{-2}$. The optimization of the string method was done using the FIRE minimizer~\cite{bitzek_structural_2006, guenole_assessment_2020}, while interpolation of the MEP was carried out with a cubic spline. This should result in an error scaling as $O(N_{images}^{-4})$~\cite{e_simplified_2007}. The location of the transition state is identified using a climbing image approach~\cite{henkelman_climbing_2000}. Finally, we check whether the converged MEP contains any intermediate minimum. If not, we preserve it as a `good' double-well potential. Those with intermediate minima can be recurrently refined by splitting into elementary double-well potentials, if these intermediate transitions are not already sampled in the dynamics.

The parameters and results of the landscape exploration are summarized in Table~\ref{table:exploration_results}. The data is split between two datasets. Dataset 1 is a coarse sampling with a long minimization period $\tau_{cg}$ and a relatively small number of independent glass metabasins $N_g$, at three preparation temperatures: $T_f = 0.509$, 0.558, 0.617. Dataset 2 is a finer and more extensive sampling, with a short minimization period $\tau_{cg}$, a large number of independent glass metabasins $N_g$ and covering four temperatures, $T_f = 0.488$, 0.509, 0.558, 0.617. The use of two data sets allowed us to validate our choices for the many parameters involved in the construction of a library of double-wells. 

\section{Tunneling states in the potential energy landscape} 

\label{sec:PEL}

\subsection{Statistics of potential energy minima}

\begin{figure*}[t]
    \includegraphics[width=\linewidth]{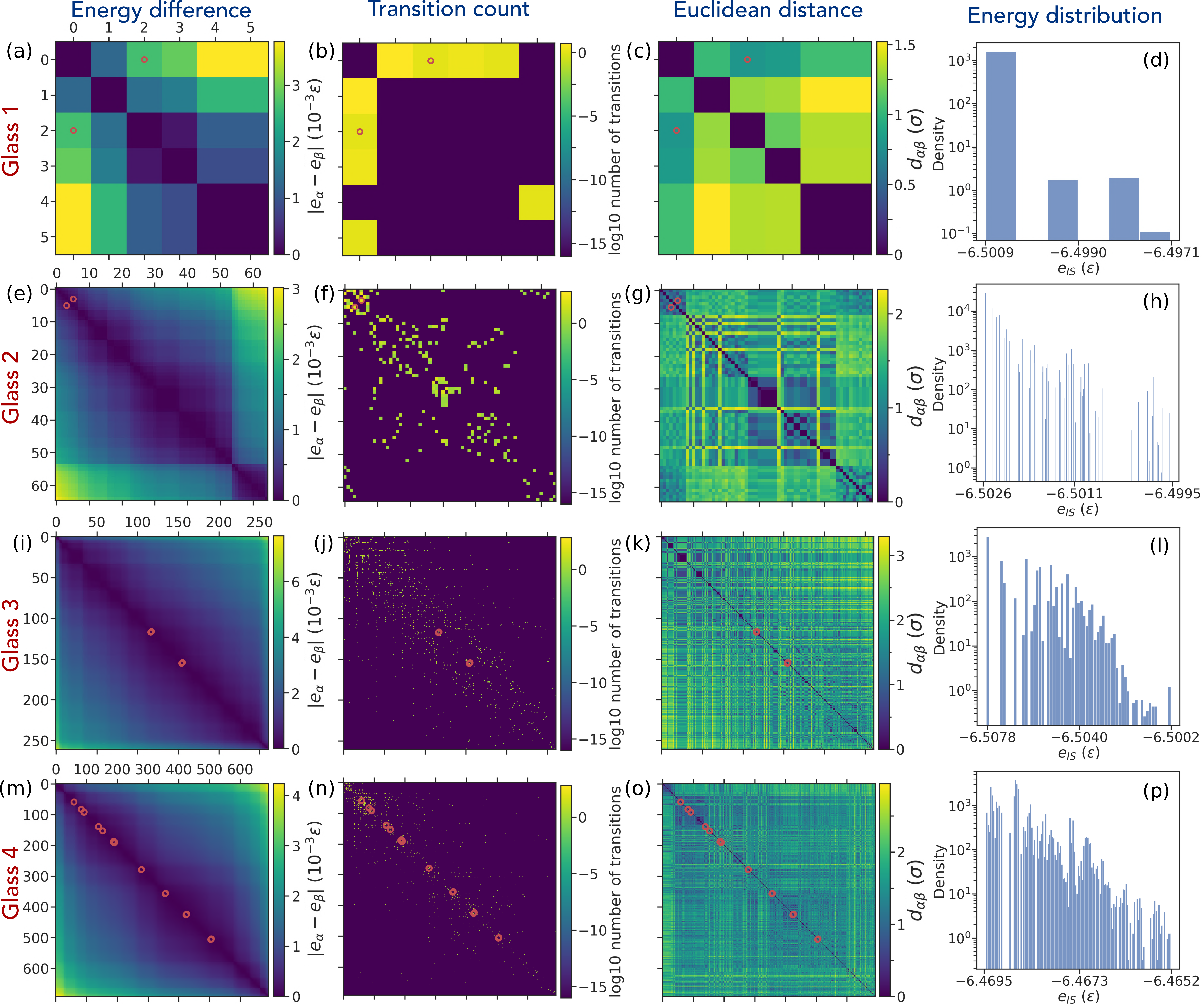}
    \caption{Heterogeneity of the potential energy landscape of four different glass metabasins labelled in \emph{red} on the left hand side. Every row of panels corresponds to a glass prepared at $T_f=0.488$ and sampled at $T_{exp}=0.4$. In each glass, distinct inherent structures (IS) are labelled with greek letters $\alpha=0,1,\ldots,n_{IS}-1$ in order of increasing energy $e_\alpha$. Each column of panels corresponds to a different observable, labeled at the top in \emph{blue}. The first three columns correspond to matrices where values are evaluated for every pair of minima in a given glass. A consistent color map (viridis) is used for all matrices, with \emph{purple} highlighting a low value and \emph{yellow} a large one. First column (a,e,i,m): matrix of IS energy difference $|e_\alpha-e_\beta|$ in units of $10^{-3} \varepsilon$. Second column (b,f,j,n): matrix $T_{\alpha\to \beta}$ counting the number of observed transitions from $\alpha$ to $\beta$, shown in a log-scale. Third column (c,g,k,o): matrix of Euclidean distance $d_{\alpha\beta}$ in units of $\sigma$. Fourth column (d,h,l,p): probability density function of IS energies. Red circles indicate the transitions that correspond to tunneling two-level systems.}
    \label{fig:matrices}
\end{figure*}

Landscape exploration of large glassy systems is an arduous task as the number of minima is expected to scale exponentially in the number of particles~\cite{sciortino_potential_2005, heuer_exploring_2008}. Equilibrium configurations prepared with swap Monte Carlo at a temperature $T_f$ ($T_g < T_f < T_{\rm mct}$) are used to start long MD runs ($\sim10^5$ steps) during which the system is quenched to a lower temperature $(T_{exp} < T_f)$. 

We can assign to each IS $\alpha$ its energy
\begin{equation}
e_\alpha = \frac1N \sum_{i<j} V_{ij}(|\mathbf{r}_i^{(\alpha)} - \mathbf{r}_j^{(\alpha)}|) \ ,
\end{equation}
and compute the Euclidean distance between a pair $(\alpha, \beta)$ of IS as
\begin{equation}
d_{\alpha\beta} = \sqrt{\dfrac1N\sum_{i=1}^N|\mathbf{r}_i^{(\alpha)} - \mathbf{r}_i^{(\beta)}|^2} \ ,
\end{equation}
where $\mathbf{r}_i^{(\alpha)}$ is the position of the $i$th particle in the $\alpha$th IS.

Fig.~\ref{fig:matrices} gives a pictorial overview of the results of the exploration process for four distinct glass metabasins at the lowest $T_f=0.488$. Each row corresponds to a glass metabasin, ordered from top to bottom by increasing number of IS sampled during exploration. From left to right, the following quantities are displayed: the matrices of the energy differences, of transitions, and of Euclidean distances between IS, and the probability density function (PDF) of the IS energies $e_{\alpha}$ observed during exploration. The IS in the matrices are ordered by increasing energy $e_\alpha$. The pairs of minima corresponding to TLS are indicated by red circles in all the matrices, they typically are found near the diagonal, which means that the two minima forming a TLS are close in energy.

The most striking aspects that emerge from these figures are (i)~the large heterogeneity of the number of IS (we find between $n_{IS}\sim6$ and $n_{IS}\sim600$ IS in a single metabasin) observed even at this low $T_f$ near the estimated $T_g$, (ii)~the fact that individual IS are clustered both in energy and in Euclidean distance, as shown, e.g., by the block structure of the $d_{\alpha\beta}$ matrix~\cite{scalliet_nature_2019,liao2019hierarchical,artiaco2020exploratory}, (iii)~the fact that the number of transitions recorded during exploration scales roughly as the number of minima, which is apparent from the sparse and linear scaling nature of non-zero elements of the transition matrix, and (iv)~the fact that low energy IS are sampled repeatedly within the metabasin and the PDF decays with increasing IS energy, as expected from a Boltzmann distribution. 

\begin{figure}[!htb]
\centering
\includegraphics[width=\linewidth]{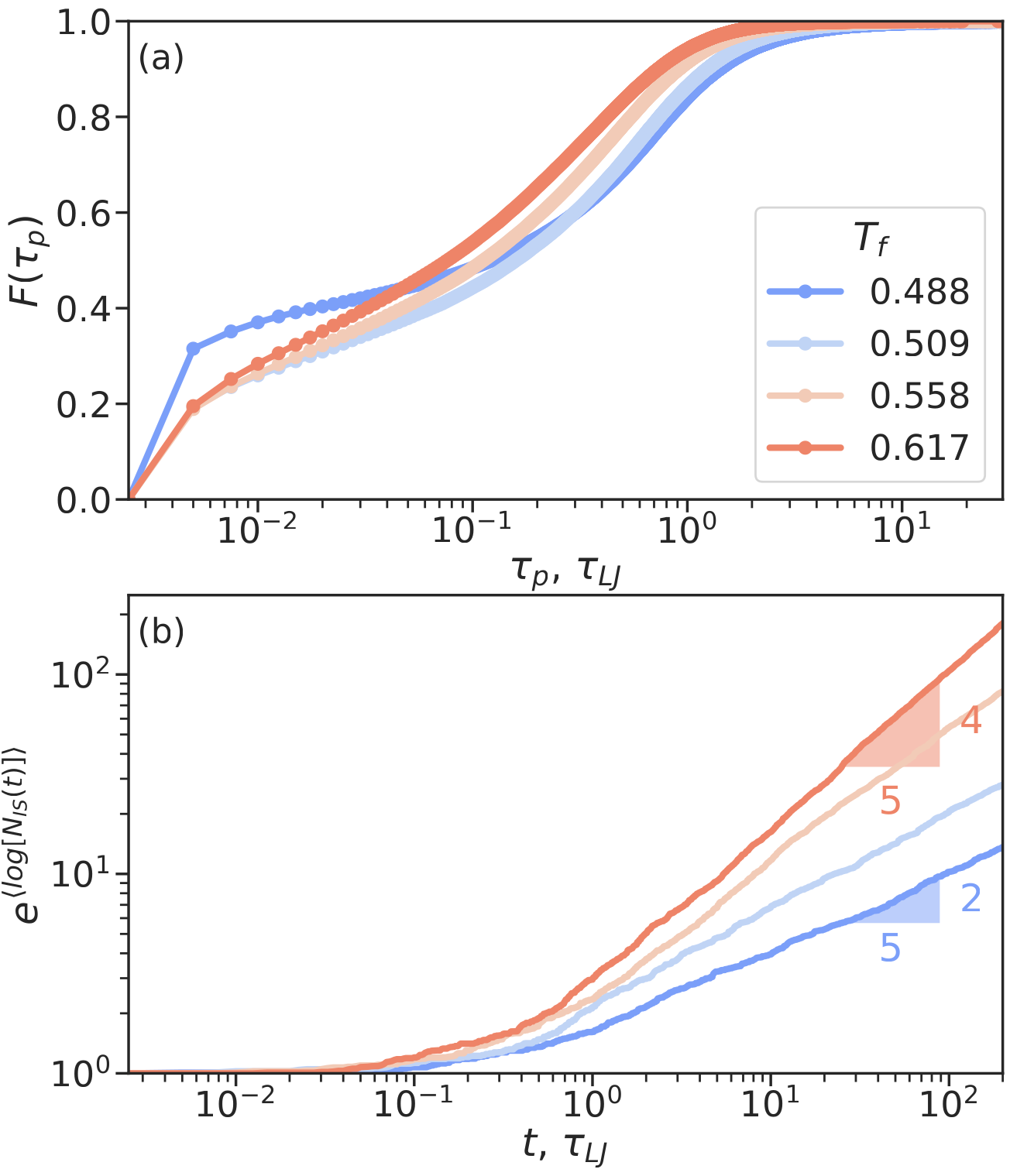}
\caption{(a) Cumulative distribution function of the persistence time $\tau_{p}$ between distinct consecutive IS in the landscape exploration when minimising every MD step at $T_{exp}=0.4$. (b) Number of newly found IS versus time, geometrically averaged over 88 distinct glasses. Colors code for the glass preparation temperature $T_f$.}
\label{fig:persist_minima}
\end{figure}

To obtain more quantitative insight, we first perform exploration runs with $\tau_{cg}=dt$, i.e. minimizing the potential energy at every MD step. The number of distinct IS per glass is counted, and a distribution of persistence times ($\tau_{p}$) is inferred from the resulting time series: here $\tau_{p}$ is the total time during which the energy minimization always ends in the same IS, before a new IS is found in the next step.  The cumulative distribution function (CDF) of  $\tau_{p}$, for different $T_f$, is shown in Fig.~\ref{fig:persist_minima}a. The shape of the CDF indicates that the distribution of $\tau_{p}$ is bimodal, with a first peak around $\tau_{p} \sim dt$ (i.e. a single MD step) and a second one around $\tau_{p} \sim 10^2 dt$. A more careful analysis of the IS time series reveals that the peak at short $\tau_p$ corresponds to processes during which the system jumps from a low-energy IS to a higher energy IS, where it stays for one or two steps, before transiting back to the low-energy IS. These processes are associated with strongly asymmetric double-well potentials, which do not give rise to TLS.

Fig.~\ref{fig:persist_minima}b shows the logarithm of the number of distinct IS obtained during exploration of a given glass metabasin, $n_{IS}(t)$, averaged over glasses, as a function of the number of steps for different $T_f$. We chose to average the logarithm of $n_{IS}$ in order to make sure that the average would not be dominated by rare metabasins with many IS. We however found that taking the logarithm of the average, i.e. $\log[N_{IS}(t)/N_g]$, yields similar results.

We find a sub-linear power-law dependence of the number of minima on the exploration time, i.e.
\begin{equation}\label{eq:ISvst}
    \langle\log(n_{IS})\rangle \propto \beta \log(t/\tau_0) \approx \log[N_{IS}(t)/N_g] \ ,
\end{equation} 
with an exponent $\beta \in [0.4,0.8]$ depending on the preparation temperatures considered. As $T_f$ decreases, the typical persistence time increases, and the total number of minima per glass, at a given time $t$, decreases. This reflects the already well-documented trend that glass metabasins tend to become much simpler in more stable glasses~\cite{scalliet_nature_2019}. 

The results of Fig.~\ref{fig:persist_minima}a suggest that one can make the exploration process more efficient by performing the energy minimization (that is computationally costly) after every $\tau_{cg}/dt$ MD steps only, instead of after every step. We have chosen $\tau_{cg}$ in our final data production runs for landscape exploration as a compromise between accuracy and efficiency. A low value of $\tau_{cg}$ results in a more accurate count of minima and transitions, with fewer cases where intuermediate minima need to be resolved. However, it can be computationally demanding to minimize too frequently as new minima appear in significant numbers only after $\sim 100-1000$ steps. The values $\tau_{cg}$ employed are provided in Table~\ref{table:exploration_results}.

We do not observe any saturation of the total number of IS, $N_{IS}(t)$, with growing exploration time $t$. As a result, we cannot perform an exhaustive search of all the IS within a glass metabasin, as can be expected for a system of mesoscopic size (1200 atoms). Yet, the average number of minima found in each glass basin, $N_{IS}/N_{g}$, is strongly depleted with increased glass stability, at each fixed $t$. We can conclude that more stable glasses have fewer IS, provided the exploration time of the metabasin exceeds $t \sim \tau_{LJ}$. The time unit $\tau_{LJ}$ seems to be close to the characteristic time for moving outside of the basin of a single IS. In the real system at \SI{1}{\kelvin} one might also expect that only a subset of minima are visited. However, in that case, exploration is dominated by tunneling through barriers and not by crossing them, as in our classical simulations. 

\subsection{Identification of double-well potentials}

\begin{figure}[t]
    \centering
    \includegraphics[width=\linewidth]{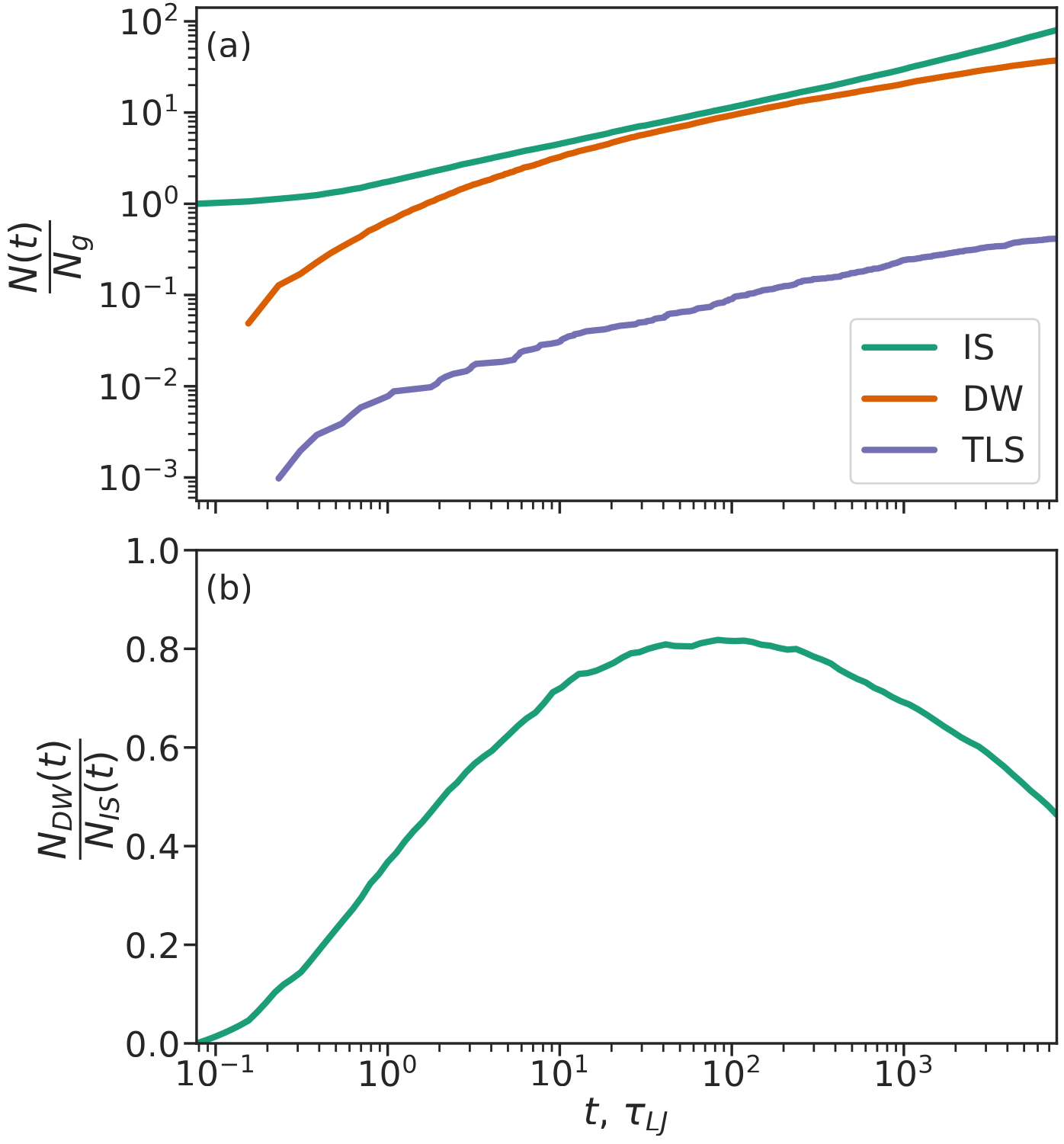}
    \caption{(a) Number of inherent structures (green), double-wells (orange) and two-level systems (purple) sampled in glasses ($T_f=0.488$) as a function of exploration time. The number of IS grows as a power-law while the growth of DW is slower at long times. TLS follow the same time evolution as DW. (b) Ratio of the number of DW to that of IS as a function of exploration time. The two panels share the horizontal axis.}
    \label{fig:trans_fpt}
\end{figure}

We now want to understand whether our landscape exploration protocol is able to identify all the relevant double well potentials. To this aim, for each DW potential, we measure the first time $t$ at which the corresponding IS pair passes all the filters to be considered a candidate transition: both IS have been included in the library, and the condition $T_{\alpha\to\beta}\geq 1$ and $T_{\beta\to\alpha}\geq 1$ are met. We then compute the number of detected DW and TLS (defined in Sec.~\ref{sec:dwp}) as a function of $t$, shown in Fig.~\ref{fig:trans_fpt}a for glasses prepared at the lowest temperature $T_f=0.488$. Similar results are obtained at all temperatures. We observe that the growth of the number of detected transitions with exploration time is slower than that of the number of IS, and is somehow intermediate between a power-law with a small exponent and a logarithmic asymptotic behavior. The growth in the number of TLS qualitatively matches that of the DW.

Assuming logarithmic behavior, comparison with Eq.~\eqref{eq:ISvst} gives $N_{DW}(t) \propto \log t \propto \log N_{IS}(t)$, which suggests that we are able to identify the proper elementary excitations of our system, as discussed in Appendix~\ref{appendix:nddwnis}, see the discussion after Eq.~\eqref{eq:KHGPSnonint}. A faster growth of $N_{DW}$ with $N_{IS}$, which remains compatible with our data, would suggest that interactions between elementary excitations play a role. This result also implies that even if we are unable to reach a proper saturation of the IS library with exploration time, we are instead able to achieve a much more satisfactory saturation of the DW library. This suggests that the new IS discovered at large times correspond to combinations of already detected excitations.

We also note that the connectivity of the explored IS, as encoded in the ratio $N_{DW}/N_{IS}$ shown in Fig.~\ref{fig:trans_fpt}b, remains of order one and has a mild dependence on the exploration time. Note that this mild time-dependence is still compatible with $N_{DW}$ being logarithmic in time, and $N_{IS}$ being a power-law with a small exponent. We provide the equivalent of Fig.~\ref{fig:trans_fpt} for the polydisperse soft sphere model in the Appendix, Fig.~\ref{fig:FtauPSS}.

\subsection{Estimation of quantum tunneling}

\label{sec:dwp}

\begin{figure}[!t]
	\includegraphics[width=.95\linewidth]{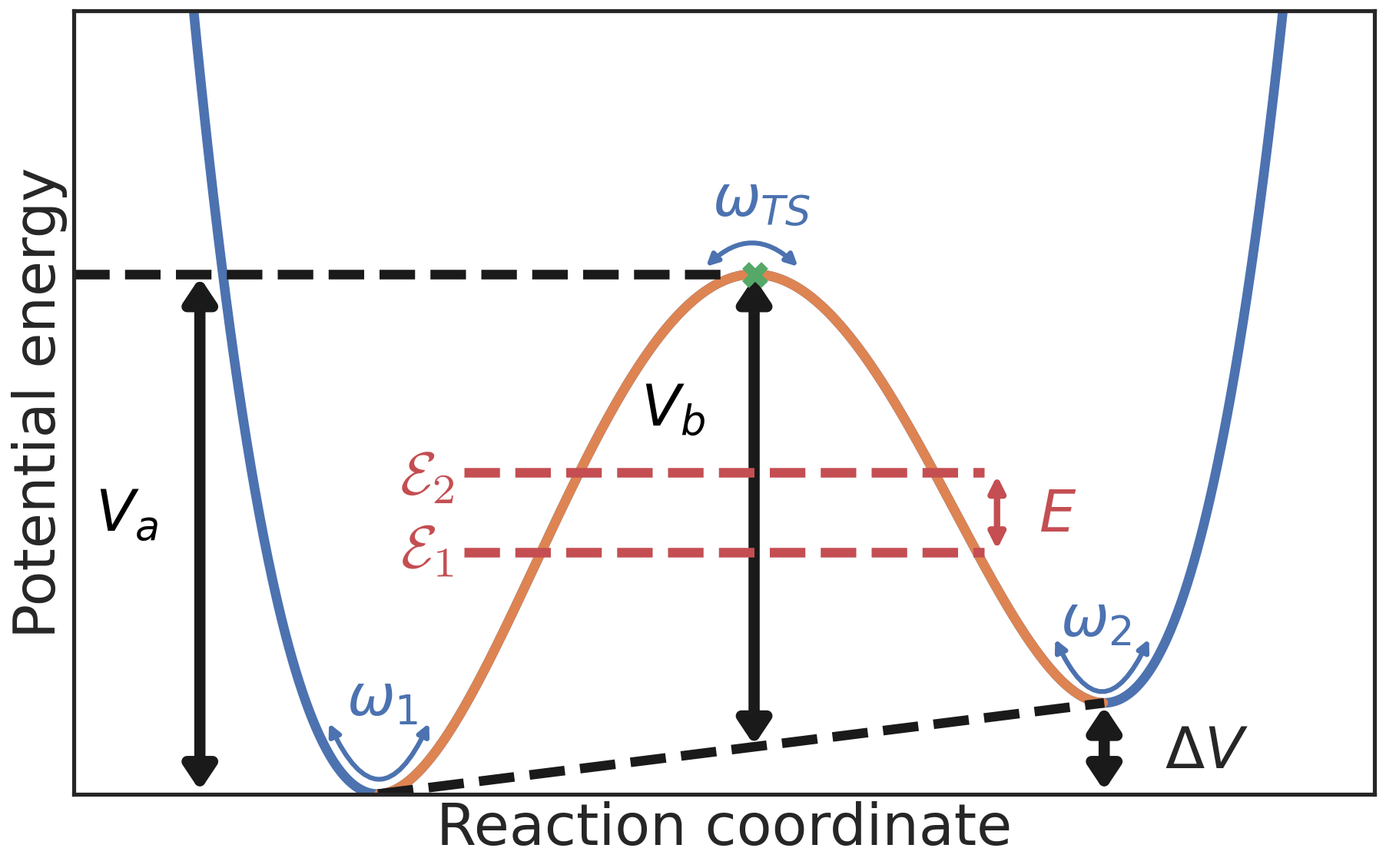}
	\caption{Schematic representation of a double-well potential illustrating the approximate MEP obtained numerically (orange) and its extrapolation beyond the two minima (blue), along with the forward barrier $V_a$, asymmetry $\Delta V$, barrier height $V_b=V_a - \Delta V/2$ (black arrows), and the curvature of the potential-energy surface at the lowest minimum $\omega_1$, highest minimum $\omega_2$, and transition state $\omega_{TS}$ (blue curved arrows). The quantum splitting $E$ is the difference between the first two quantum energy levels ${\varepsilon}_1$ and ${\varepsilon}_2$ (dashed red).}
	\label{fig:scheme}
\end{figure}

Two-level systems within our glassy models correspond to tunneling DW potentials with a low quantum splitting. The temperature scale below which quantum effects are important is $T_Q \sim$~\SI{1}{\kelvin} in experiments and can be obtained from comparing the interparticle distance with the thermal wavelength, leading to:
\begin{equation}
    T_Q = \frac{2 \pi \hbar^2}{m \sigma^2 k_B} \ .
\end{equation}
The MEP obtained via a converged optimization of the string method, with no intermediates, results in a 1D potential energy profile, as schematized in Fig.~\ref{fig:scheme}. By convention, the lowest energy minimum is taken to be the one on the left. From this simplified 1D potential, several quantities can be defined: the forward barrier, $V_a$, is the energy difference between the transition state and the lowest energy minimum; the asymmetry, $\Delta V$, is the difference in energy between the two minima, and the barrier height can be obtained from these two quantities: $V_b = V_a - \frac{1}{2}\Delta V$.

As a first approximation, we reduce the tunneling problem to the effective one dimensional potential of the MEP. Considering a normalized reaction coordinate $\xi \in [0, 1]$ which corresponds to the fraction of the Euclidean distance $d$ along the minimum energy path, we can write the resulting Schr\"odinger equation as
\begin{equation}
	-\dfrac{\hbar^2}{2 m d^2 \epsilon} \partial_{\xi}^2 \psi(\xi) + V(\xi)\psi(\xi) = \varepsilon \psi(\xi) \ .
\end{equation}
An interpolated MEP obtained from the NEB calculation is discretized by $d\xi=d/N_\xi$, where $N_\xi=2000$, using a cubic spline. The Laplacian is evaluated with a 5-point finite difference stencil. We use $\hbar=1$ to evaluate the quantum properties of the DW. This leads to a dimensionless effective mass parameter $\tilde{m}$ which controls the ``quantumness" of the problem, as first introduced by Vineyard~\cite{vineyard_frequency_1957},
\begin{equation}
\tilde{m} = m\dfrac{\epsilon \sigma^2}{\hbar^2} \ .
\end{equation}

The MEP and the effective mass $\tilde{m}$ define a 1D potential for the DW excitation. We assume that the tunneling problem can be treated in a 1D approximation when the relevant classical path, the MEP, is nearly independent from the others~\cite{demichelis_properties_1999}. We have calculated the eigenvalues and eigenvectors of the Hessian along the MEP to check that orthogonally to the MEP, the dynamics is harmonic and independent from the reaction coordinate. The results are summarised in the Appendix and in Fig.~\ref{fig:eigMEP}.

The $\tilde{m}$ parameter depends on the choice of units. Nevertheless, it can be tuned over a wide range of values ($10^2-10^5$) without changing the qualitative behavior of the quantum splitting distribution obtained from our simulations~\cite{khomenko_depletion_2020}. In this case, for the two sets of units detailed in Sec.~\ref{sec:ivb}, namely argon (\ce{Ar}) and nickel-phosphorous (\ce{NiP}), the value of $\tilde{m}$ is 1200 and 5250, respectively. The corresponding values of $k_B T_Q$ are $0.005 \epsilon$ (for \ce{Ar}) and $0.0012\epsilon$ (for \ce{NiP}). 

The five smallest eigenvalues $\varepsilon_1, ...,\varepsilon_5$ are then evaluated using ARPACK~\cite{lehoucq_arpack_1998}. The splitting is given by the first two energy levels, $E = {\varepsilon}_2 - {\varepsilon}_1$. We consider as tunneling TLS all the filtered and tunneling DW that have a splitting $E < T_Q$. Following our notations, their total number is $N_{DW}(T_Q)$.

\begin{figure*}[t]
\includegraphics[width=\linewidth]{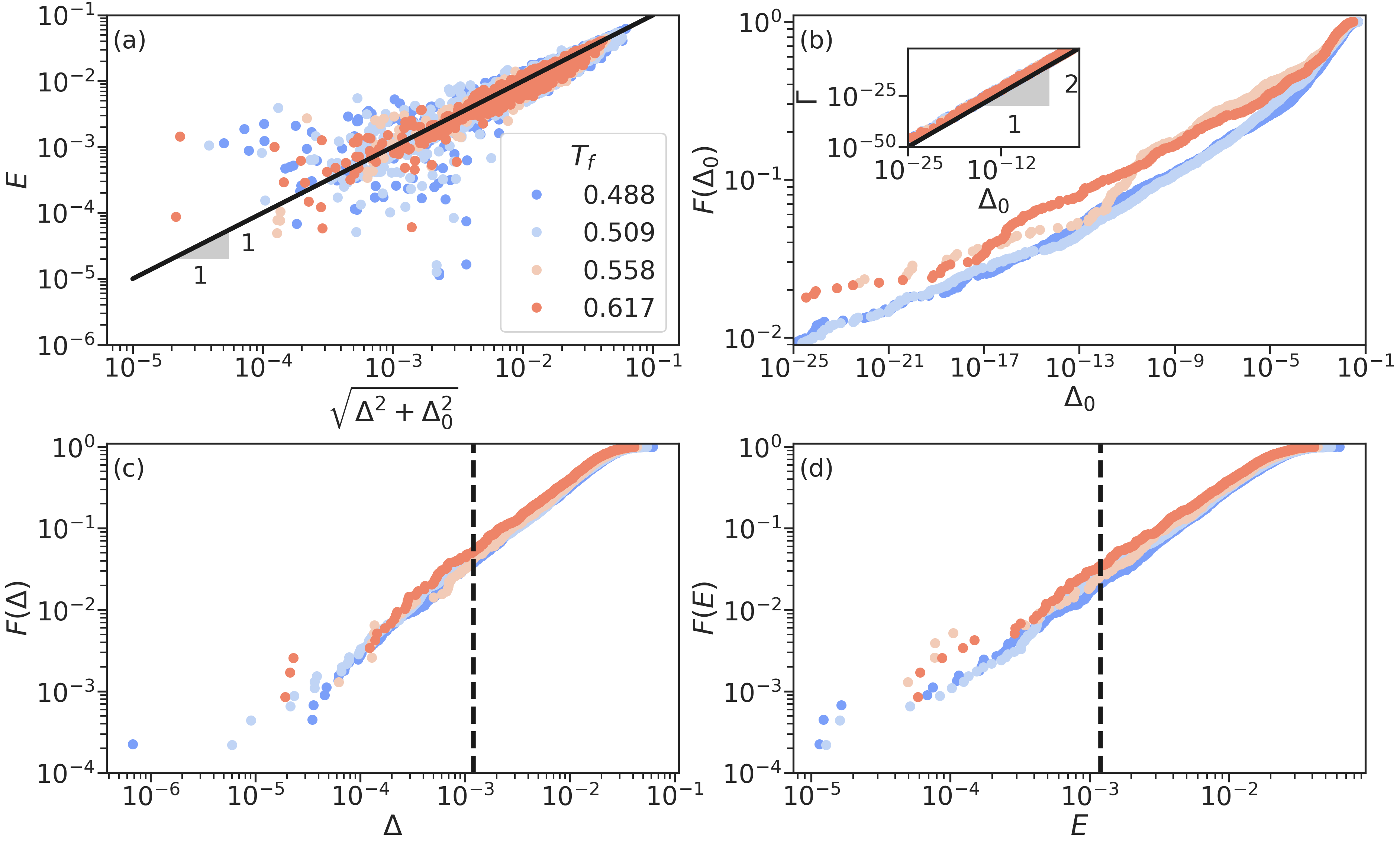}
	\caption{(a) Scatter plot of the quantum splitting $E$ versus $\sqrt{\Delta^2 + \Delta_0^2}$. Colors code for different preparation temperatures $T_f$. (b) Cumulative distribution function of the tunneling matrix element $\Delta_0$. Inset: decay rate $\Gamma$ versus $\Delta_0$, following a $\Gamma \propto \Delta_0^2$ scaling (line). Cumulative distribution functions of (c) the diagonal splitting $\Delta$ and (d) the quantum splitting $E$. Dashed lines indicate $T_Q = 0.0012$. The linear behavior in (c,d) below $T_Q$ directly validates the hypothesis of the TLS model. Data in all panels were calculated using $\tilde{m}=5250$, corresponding to \ce{NiP} units. }
	\label{fig:wkb}
	\label{fig:n0cdf}
      \end{figure*}

We can also attempt to map the DW profile into an actual two-level system.
In the standard TLS model, the Hamiltonian of a single tunneling state takes the form
 \begin{equation}
	H = \frac{1}{2}\begin{pmatrix}
		\Delta & \Delta_0\\
		\Delta_0 & -\Delta
	\end{pmatrix}
\end{equation}
in the localized representation. The diagonal splitting of the TLS can be estimated by
\begin{equation}
	\Delta = \Delta V + \hbar \frac{\omega_2 - \omega_1}{2} \ ,
\end{equation}
where $\omega_1$ and $\omega_2$ are the characteristic frequencies of the two minima of the double-well potential, see Fig.~\ref{fig:scheme}. The decay rate of the TLS can be evaluated in the Wentzel–Kramers–Brillouin (WKB) approximation as
\begin{equation}
\begin{split}
    \Gamma &= \frac{1}{m} \left[ \int_{0}^{a} \frac{dx}{p(x)} \right]^{-1} \exp{ \left[ -\frac{2}{\hbar} \int_{a}^{b} |p(x)| dx \right]}, \\
    p(x)&=\sqrt{2 m ({\varepsilon}_2 - V(x))} \ .
\end{split}
\end{equation}
The tunneling matrix element $\Delta_0$ can be obtained from the effective one-dimensional potential via the WKB approximation as
\begin{equation}
	\Delta_0 \approx \bar{{\varepsilon}} \exp{ \left[ -\frac{1}{\hbar} \int_{a}^{b} |p(x)| dx \right]}, \quad \bar{{\varepsilon}} = \frac{{\varepsilon}_1 + {\varepsilon}_2}{2} \ .
\end{equation}
      
The integration limits $a$ and $b$ correspond to values of the effective reaction coordinate $x$ where a certain energy level crosses the double-well potential curve at either side of the barrier. This level is ${\varepsilon}_2$ for the calculation of $\Gamma$, and $\bar{{\varepsilon}}$ for $\Delta_0$.       

As such, the quantities $\Delta$ and $\Delta_0$ are expected to approximately obey the relationship $E \approx \sqrt{\Delta^2 + \Delta_0^2}$. The quantum splitting $E$ is plotted against $\sqrt{\Delta^2 + \Delta_0^2}$ in Fig.~\ref{fig:wkb}a, showing the expected correlation at all temperatures. Small deviations from $E\approx\sqrt{\Delta^2 + \Delta_0^2}$ are observed at low splittings $E$. These deviations could appear due to the breakdown in the various approximations that are made to calculate $E,\Delta$ and $\Delta_0$. The magnitude of the deviations becomes important as $E\sim1/N$ suggesting that finite-size effects could influence the absolute magnitude of very low splittings. 

\begin{figure*}[t]
	\centering	 
	\includegraphics[width=\linewidth]{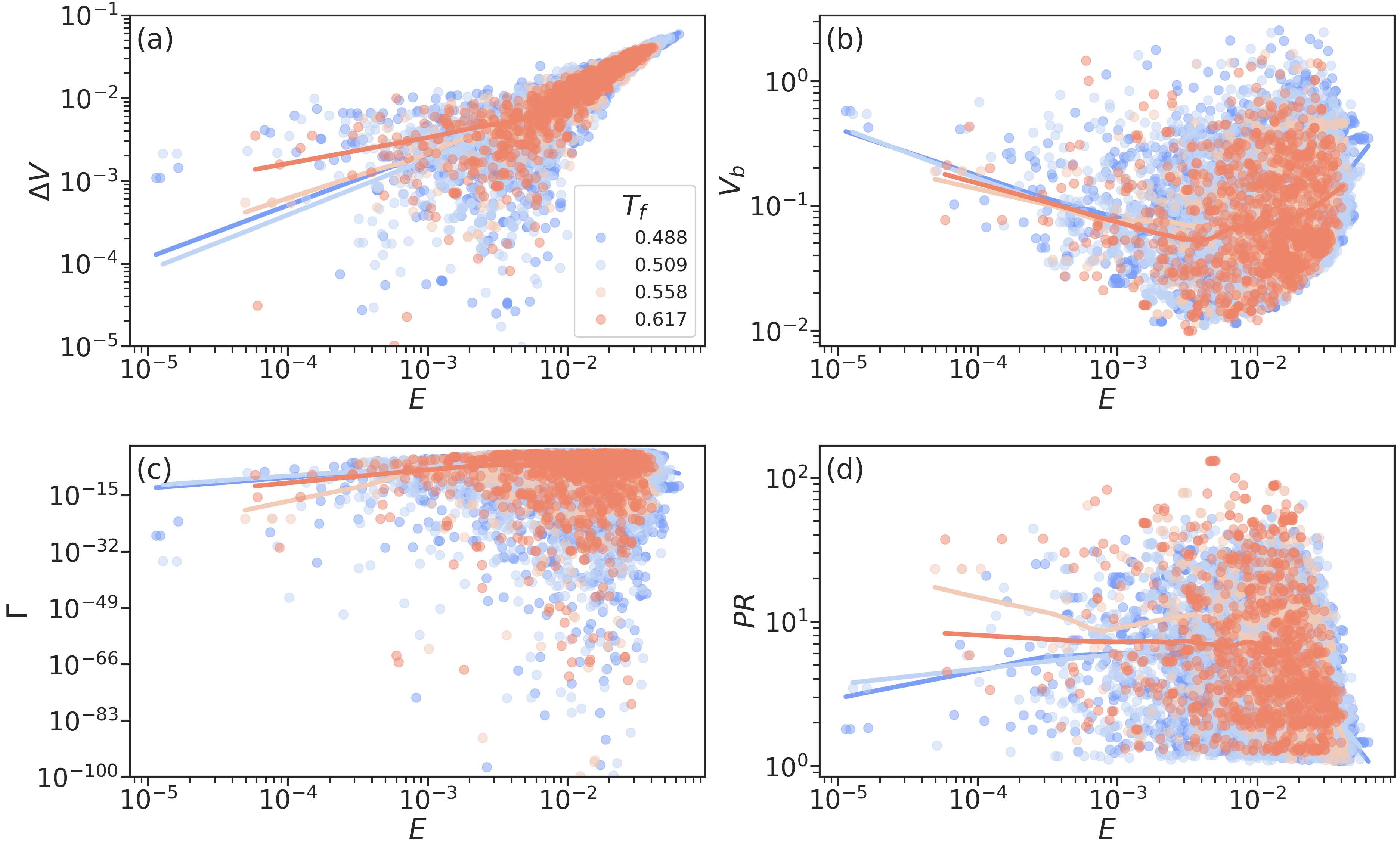}
	\caption{Statistics of tunneling double-well potentials versus the estimated quantum splitting $E$: (a) asymmetry $\Delta V$, (b) barrier height $V_b$, (c) tunneling decay rate $\Gamma$, and (d) participation ratio $PR$. All values were calculated with a reduced mass $\tilde{m}=5250$, corresponding to \ce{NiP} units. Colors code for different preparation temperatures $T_f$. Solid lines correspond to average values at a given $E$ obtained via locally weighted regression.}
	\label{fig:splitpannel}
\end{figure*}

Fig.~\ref{fig:wkb}b shows the cumulative probability distribution function $F(\Delta_0)$ of the tunneling matrix element $\Delta_0$.
The TLS model predicts a flat PDF for $\log \Delta_0$, which corresponds to $F(\Delta_0) \sim \log \Delta_0$ at small $\Delta_0$. Our data instead suggest that $F(\Delta_0) \sim \Delta_0^{\varphi}$ with a very small exponent $\varphi\approx 0.1$, corresponding to a PDF $p(\Delta_0)\sim \Delta_0^{\varphi-1}$. Such behavior was discussed in more detail elsewhere~\cite{frossati1977spectrum,phillips_two-level_1987}.

The decay rate $\Gamma$ is also expected to approximately obey $\Gamma \propto\Delta_0^2E$ due to Landau-Zener tunneling ~\cite{jaksic_landauzener_1993}, which is seen in the inset of Fig.~\ref{fig:wkb}b. 
Finally, the CDFs of $\Delta$ and $E$ are shown in Fig.~\ref{fig:wkb}c and Fig.~\ref{fig:wkb}d, respectively. Below $T_Q$, both the CDFs of $\Delta$ and $E$ are linear, as expected in the TLS model, and change very little with $T_f$.

The ratio of tunneling TLS to generic double-wells, $N_{DW}(T_Q)/N_{DW}$, seems to be roughly  constant $\sim0.3\%$, at all temperatures. Because the function $N_{DW}(E)/N_{DW}$ converges to a stable finite limit when $N_{DW}\to\infty$, the ratio $N_{DW}(T_Q)/N_{DW}$ is also stable in both datasets presented in Table~\ref{table:exploration_results}, indicating that it is not particularly sensitive to the details of landscape exploration.

\section{Microscopic properties of two-level systems}

\label{sec:TLS}

\subsection{Statistical properties of two-level systems}

The data from the extensive exploration of the energy landscape is summarized in Fig.~\ref{fig:splitpannel}, which presents the statistics of different observables of tunneling DW potentials in relation to their estimated quantum splitting $E$ and the preparation temperature $T_f$ of the glass. The DW asymmetry, $\Delta V$, is shown in Fig.~\ref{fig:splitpannel}a where one can notice a threshold value $\Delta V \sim 10^{-2}$, below which the distribution of quantum splittings becomes much broader. Asymmetry between energy minima seems to be on average higher for higher preparation temperature $T_f$. 

The statistics of the energy barrier, $V_b$, versus $E$ are shown in Fig.~\ref{fig:splitpannel}b. Our dataset probes a range of two orders of magnitude in barrier heights between sampled tunneling DW. In general, DW with a low splitting $E$ have a relatively high barrier $V_b$. DW sampled in lower $T_f$ glasses typically have a higher barrier than those found in glasses prepared at high $T_f$.

The statistics of the tunneling decay rate $\Gamma$ ($\propto \Delta_0^2 \times E$) are shown in Fig.~\ref{fig:splitpannel}c. The distribution of tunneling decay rates $\Gamma$ becomes narrower as the tunnel splitting $E$ decreases. On average the decay rate $\Gamma$ decreases with the quantum splitting $E$. Roughly one third of the sampled tunneling DW decay in more than \SI{1}{hour}.

Our numerical study provides information on the microscopic properties of the transition sampled. From a structural point of view, the degree of localization can be evaluated from the atomic displacement $d^i_{\alpha\beta}=|\mathbf{r}_i^{(\alpha)} - \mathbf{r}_i^{(\beta)}|$ of atom $i$ between the two IS $\alpha$ and $\beta$, by calculating the participation ratio (PR)
\begin{equation}
\label{eq:PR}
PR = \dfrac{\left[ \sum_{i} (d^i_{\alpha\beta})^2 \right]^2}{\sum_{i}{( d^i_{\alpha\beta}})^4} \ ,
\end{equation}
defined such that $1 \leq PR \leq N$. The participation ratio indicates the number of atoms involved in a transition. 

The statistics of $PR$ versus $E$ are shown in Fig.~\ref{fig:splitpannel}d. Most of the displacements observed in our database are localized, and they have an average participation ratio $PR\sim8$. Even the most delocalized DW that pass the filtering procedure do not have a $PR$ much larger than $100$. We see from Fig.~\ref{fig:splitpannel}d that on average, the $PR$ decreases with decreasing $T_f$. This can be rationalized in terms of displacements that tend to be more local in more stable glasses. 

However, examples of delocalized DW ($PR\sim100$) with a low $E$ do exist, but they are found only in glasses prepared at high $T_f \simeq T_{\rm mct}$. Such delocalized excitations correspond to very small individual displacements and have a small barrier. This is a similar pattern to what was observed also in the atomic tunneling of complex crystalline defects such as kinks in metallic \ce{Cu}~\cite{vegge2001}.

\begin{figure*}[t]
	\centering	
	\includegraphics[width=\linewidth]{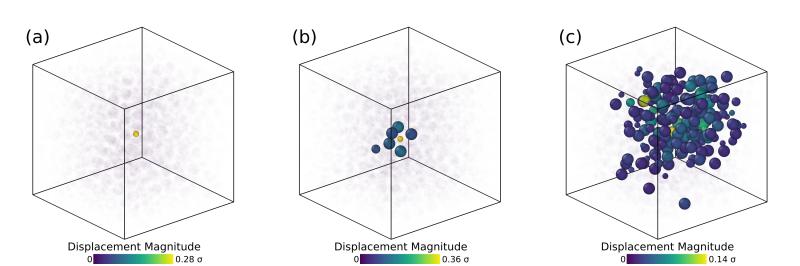}
	\caption{Visualization of atomic displacements between the two minima forming a double-well potential (DW) with low quantum splitting. Only the particles that displace the most are shown, and the rest are made transparent and faded in the background for clarity. (a)~A localized two-level system in which a single particle moves in an almost frozen structure, $PR\sim1$. (b)~A typical two-level system, with $PR\sim7$. (c)~A delocalized two-level system with $PR\sim83$; these cases are rare and mostly occur in less stable glasses $T_f = 0.617$.} 
	\label{fig:tls}
\end{figure*}

We have also looked at the scatter plot of Euclidean distances versus the quantum splitting which is shown in Fig~\ref{fig:dvsE} in the Appendix. The patterns of real-space atomic displacements between the configurations of the two IS forming a DW potential are illustrated in Fig.~\ref{fig:tls}. Atoms are shown with a radius proportional to their interaction range $\sigma$ and colored by $d^i_{\alpha\beta}$, the magnitude of their displacement between the two minima. In order to ease visualization only the particles that displace the most are shown, with the rest faded into the background for clarity. For clarity, only the first particles up to PR, to the nearest integer, are highlighted. In general, displacements in our database of DW are highly localized and involve only several atoms. We nevertheless can distinguish three classes of TLS, similar to a recent study on amorphous \ce{Si}~\cite{levesque_internal_2022}. The most localized DW potentials have a $PR \sim 1$ such as the one shown in Fig.~\ref{fig:tls}a. These would correspond to defect hopping within the material. A more typical TLS is shown in Fig.~\ref{fig:tls}b, it has a $PR\sim7$. It is still one atom that moves significantly, but this triggers the rearrangement of nearest neighbours as well. Although very rare, we also do find delocalized excitations, see Fig~\ref{fig:tls}c, where nearly 10\% of the particles move a relatively small distance throughout the whole structure. These tend to occur in less stable, higher $T_f \simeq T_{\rm mct}\simeq 0.62$ glasses.

\subsection{Depletion of tunneling two-level systems with increasing glass stability}
\label{sec:ivb}

\begin{figure}[!ht]
	\includegraphics[width=\linewidth]{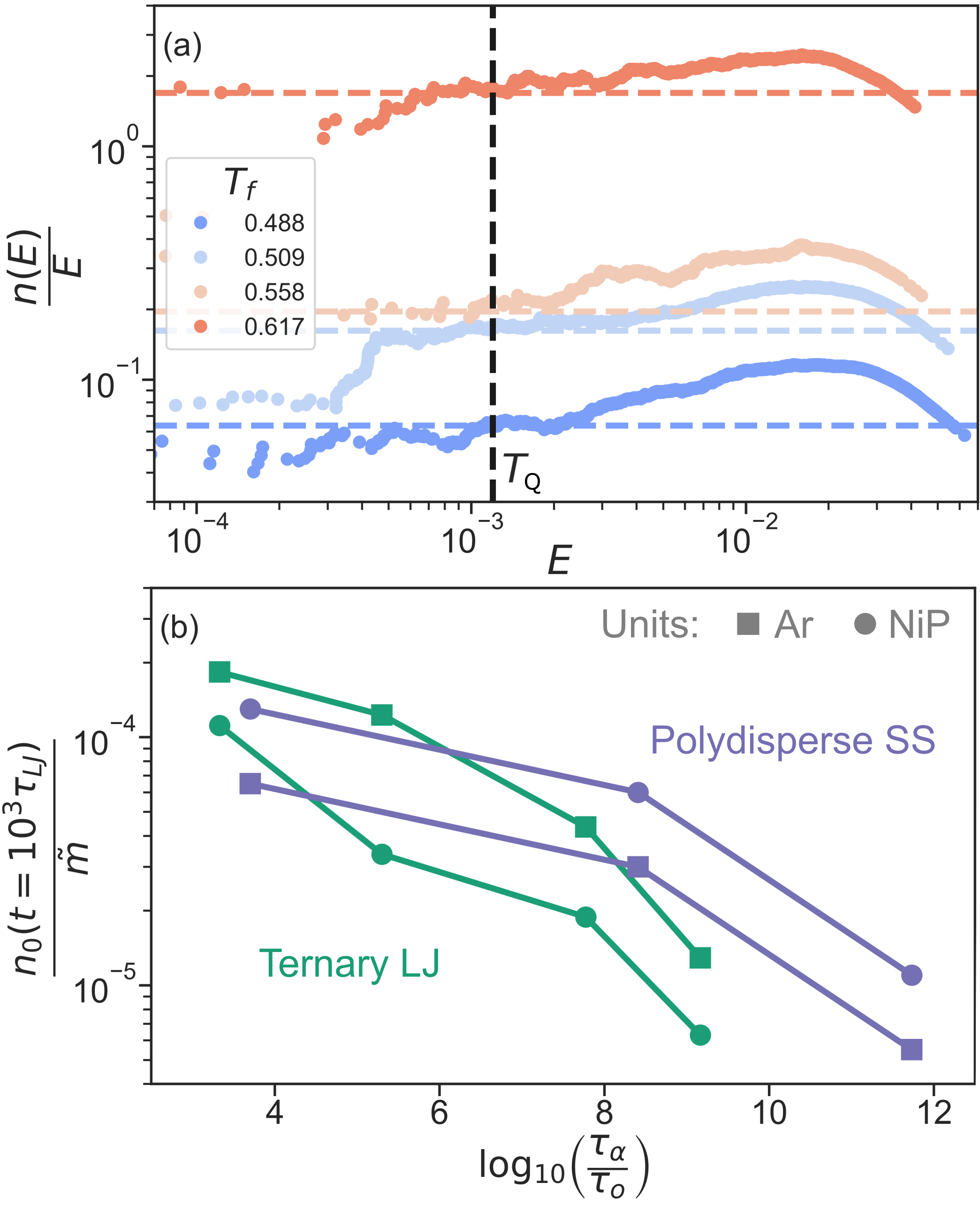}
	\caption{(a) Cumulative distribution of the quantum splitting $E$ for all tunneling double-well potentials, as a function of glass stability $T_f$ ($\tilde{m}=5250$). Plateau values indicate $n_0$. (b) Density of two-level systems $n_0$ in ternary Lennard-Jones (TLJ) and polydisperse soft-sphere (PSS) glasses as a function of stability, as encoded in $\tau_{\alpha}(T_f)/\tau_o$. Data for two sets of units: \ce{NiP} (circle, $\tilde{m}=5250$), and Ar (square, $\tilde{m}=1200$). PSS data from Ref.~\cite{khomenko_depletion_2020}. To compare the models we have fixed the exploration time $t=10^3 \tau_{LJ}$ to estimate $n_0$.}
	\label{fig:n0tau}
\end{figure}

The standard TLS model predicts a plateau $n(E)/E \to n_0$ as $E\to0$, where $n(E)$ is the cumulative distribution of quantum splittings. The results obtained from our energy landscape exploration of the ternary LJ model are shown in Fig.~\ref{fig:n0tau}a. In agreement with the tunneling model, we observe a plateau for $E < T_Q$ (vertical dashed) at all $T_f$, and a peak at $E\sim5\times10^{-2}$. In addition to the low-$E$ plateau behavior observed at each temperature, we identify a clear depletion of $n(E)/E$ as $T_f$ decreases. This agrees with the previous computational estimation of the TLS density as a function of glass stability in polydisperse soft-spheres (PSS)~\cite{khomenko_depletion_2020}. 

In order to compare the TLJ data with the PSS model we have plotted $n_0/\tilde{m}$ versus the stability of the glass, as encoded in the relaxation time ratio $\log (\tau_\alpha(T_f)/\tau_o )$, in Fig.~\ref{fig:n0tau}b. In this representation, glass stability increases from left to right. The relaxation time $\tau_\alpha(T_f)$ is either directly measured, or estimated by extrapolating the data to lower temperature using a parabolic law, which was shown to perform well such  extrapolation~\cite{AGtest}. The $n_0$ values for PSS are from Ref.~\cite{khomenko_depletion_2020}, while we have refined our $\tau_\alpha$-extrapolation using the latest data obtained via long MD simulations of PSS~\cite{scalliet_thirty_2022}.  
The range of glass stabilities explored in the ternary model is roughly two orders of magnitude smaller than in PSS. This is due to the particle-swap algorithm thermalizing more efficiently continuously polydisperse mixtures compared to ternary mixtures. To make a meaningful comparison of the defects in both models, we have estimated the TLS density $n_0$ after a fixed exploration time of $t=10^3\tau_{LJ}$. This is the longest exploration time that can be probed from our simulations, in both models, at all $T_f$. This restriction has reduced the range of $n_0$ values showed previously for the PSS model~\cite{khomenko_depletion_2020}. Nevertheless, at this fixed exploration time, the two models show a similar depletion of two-level systems, with the TLJ model showing a slightly steeper depletion. As evidenced from Table~\ref{table:exploration_results}, this depletion of tunneling defects seems to be driven by a reduction in the number of IS available in a glass metabasin as $T_f$ decreases. The curves are shown for different values of $\tilde{m}$ to probe quasi-universality by emulating different material properties within the same model. This confirms that the precise value of $\tilde{m}$ does not affect our main conclusion. 

The choices of $\tilde{m}$ prompt a short discussion on the units. Two sets of physical units have been investigated in order to make a connection to experimental observations. The first one corresponds to physical units that mimic pairwise interactions between argon (Ar) atoms~\cite{demichelis_properties_1999}, and has the following parameters: $\sigma=3.405 \times 10^{-10}$ \SI{}{\meter}, $\epsilon/k_B=$ \SI{125.2}{\kelvin} and a mass of $m=6.634 \times 10^{-26}$ \SI{}{\kilo \gram}. The unit of time is $\tau_{LJ}=$\SI{2.1}{\pico \second}. In these units, the numerically estimated glass transition temperature is $T_g \approx$ \SI{59}{\kelvin}, $T_Q \approx$ \SI{0.658}{\kelvin} and $\tilde{m} \approx 1200$. For the preparation temperatures $T_f=0.488$, $0.509$, $0.558$ and $0.617$ one gets $n_0^{sim} \sim 0.04$, $0.09$, $0.14$ and $0.674$ $\epsilon^{-1} \sigma^{-3}$ which corresponds to $n_0^{exp} \sim 4.05\times10^{47}$, $1.19\times10^{48}$, $2.61\times10^{48}$ and $8.95\times10^{48}$ \SI{}{\per \joule \per \cubic \meter}.

The second set of units, used to mimic the average pairwise interactions in a \ce{NiP} glass~\cite{weber_local_1985, reinisch_how_2004}, consists of the following parameters: $\sigma=2.21\times 10^{-10}$ \SI{}{\meter}, $\epsilon/k_B =$ \SI{934}{\kelvin} and an average mass per particle of $m=9.266 \times 10^{-26}$ \SI{}{\kilo \gram}. The unit of time is $\tau_{LJ}=$\SI{2.5}{\pico \second}. In these units, the glass transition temperature is $T_g \approx$ \SI{438}{\kelvin}, $T_Q \approx$ \SI{1.118}{\kelvin} and $\tilde{m} \approx 5250$. The defect density is $n_0^{sim} \sim 0.08$, $0.19$, $0.23$ and $1.84$ $\epsilon^{-1} \sigma^{-3}$ which corresponds to $n_0^{exp} \sim 4.24\times10^{47}$, $1.11\times10^{48}$, $1.53\times10^{48}$ and $1.17\times10^{49}$ \SI{}{\per \joule \per \cubic \meter}. 

We find a modest variation in the absolute value of the tunneling defect density with $\tilde{m}$. The difference between the defect density of the two models (PSS and TLJ) at the same glass stability is similar in magnitude to the difference between the defect density calculated for different values of $\tilde{m}$ and the data is insufficient to establish any clearer pattern. 

A reduction of the number of defects could be expected in the TLJ model due to the change from a polydisperse system in which every particle is different, to one with only three types of indistinguishable particles (A,B,C). An additional influence could come from the attractive interactions, absent in the PSS model, which have been recently suggested to have an important effect on the elastic properties of glassy solids. In particular, a reduction of the density of quasi-localized modes (QLM) was reported, as the attractive part, or ``stickiness" of the pair-potential increases~\cite{gonzalez-lopez_mechanical_2021}. The decrease in QLM density can be up to an order of magnitude and was compared to the reduction in defects observed during thermal annealing. We defer the discussion of the behavior of QLM in these models and their potential relationship to $n_0$ to the next section.

\subsection{Comparison to quasi-localized harmonic modes}

To begin evaluating the vibrational modes of the TLJ glass configurations, we first estimate their elastic moduli at different $T_f$ from athermal quasistatic deformation simulations. With an initial IS obtained through energy minimization from an equilibrium configuration, a suitable cycle of small deformations is applied, each deformation being followed by an energy minimization, until a stress-strain curve is obtained. The shear modulus $G$ is obtained from athermal quasi-static shear simulations (AQS) under Lees-Edwards boundary conditions, while the bulk modulus $K$ is obtained from quasistatic deformation with periodic boundary conditions and a hydrostatic strain. The elastic moduli are obtained from the elastic regimes of the corresponding stress-strain curves,
\begin{figure}[t]
	\includegraphics[width=\linewidth]{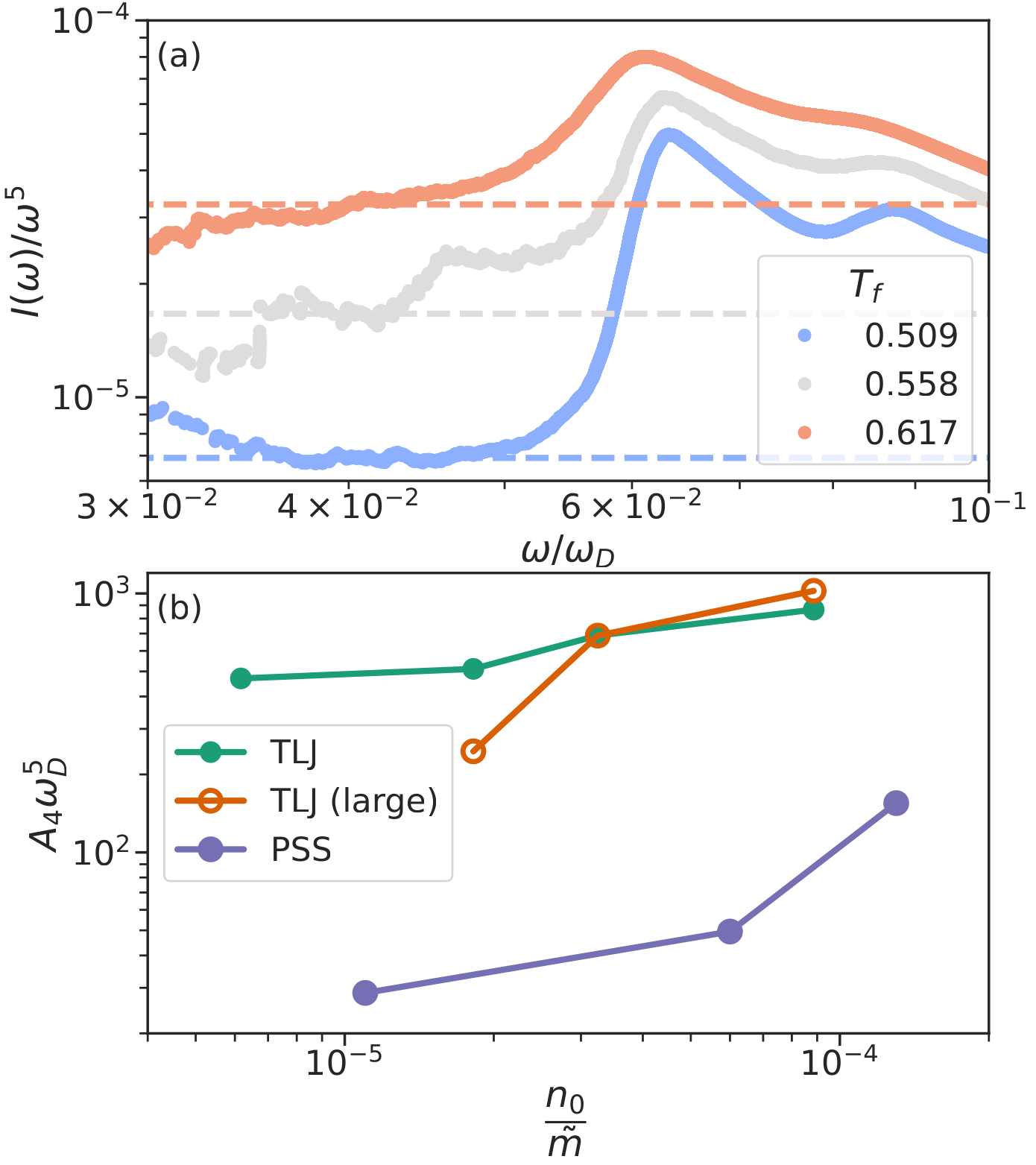}
	\caption{
	(a) Integrated density of states, $I(\omega)/\omega^5$ in large $N = 12000$ ternary Lennard-Jones (TLJ) glasses prepared at different $T_f$. Dashed lines represent the $A_4/5$ fits of $I(\omega)/\omega^5$ just below $\omega_t$. (b) $A_4$ versus $n_0$ for $N=1200$ (green) and larger $N=12000$ (orange) TLJ glasses (orange circles) and the polydisperse soft sphere (PSS) configurations from Ref.~\cite{khomenko_depletion_2020} (purple). The A$_4$ and $\omega_D$ values for the PSS model are from Ref.~\cite{wang_low-frequency_2019}.}
	\label{fig:qlmtls}
\end{figure}
\begin{equation}
    G = \dfrac{\sigma_{xy}}{\gamma_{xy}} \ , \qquad
    K = - V \dfrac{\Delta p}{\Delta V} \ .
\end{equation}

Once the moduli have been determined, we calculate the transverse $c_t = \sqrt{G / (m \rho)}$ and longitudinal $c_l= \sqrt{(K+\dfrac{4G}{3})/(m\rho)}$ sound velocities, and from them the frequency $\omega_t$ of the first phonon,
\begin{equation}
\omega_t = \dfrac{2\pi}{L} c_t \ ,
\end{equation}
and the Debye frequency,
\begin{equation}
\omega_D = \left( \dfrac{18 \pi^2 \rho}{2 c_t^{-3} + c_l^{-3}} \right)^{\frac{1}{3}} \ .
\end{equation}
We have calculated and diagonalized the dynamical matrix to obtain the vibrational modes and their frequencies. This was done for the ensemble of structures in Table~\ref{table:exploration_results}, as well as for larger 12000-atom TLJ configurations at $T_f=0.509$, 0.558 and $0.617$, using ensembles of ~1000 configurations at each $T_f$. Non-phononic vibrational modes can be seen below the first phonon peak at $\omega_t$, when the model size is small enough to avoid significant hybridization with phonons. It is typically assumed that the density of states $D(\omega)$ scales as $\omega^4$ in the low-frequency limit~\cite{kapteijns_universal_2018}.

In Fig.~\ref{fig:qlmtls}a we show the integrated density of states $I(\omega)$, scaled by the expected $\omega^5$ for the larger 12000-atom models at each $T_f$. Provided $D(\omega)\propto A_4 \omega^4$, we would expect a plateau in $I(\omega) / \omega^5$ just below $\omega_t$. We find this to be the case, however, the plateau (dashed) persists only over a limited range of frequency, below which the scaling seems to change, in agreement with recent results on the behavior of very-low frequency modes in the standard KA model~\cite{wang_density_2022}. The low-frequency behaviour of $D(\omega)$ is currently under intense scrutiny, and this discussion would require much more statistics than we currently possess. Our goal is simpler, as we only wish to understand whether the evolution of localized harmonic modes in $D(\omega)$ and of TLS are strongly correlated. 

Moving on to the quantitative relationship between QLM and TLS, we report in Fig~\ref{fig:qlmtls}b the adimensional quantity $A_4 \omega_D^5$ versus the density of TLS obtained with \ce{NiP} units. Again the TLS density $n_0$ is compared after a fixed exploration time of $10^3\tau_{LJ}$ for both models. Since both quantities decrease with decreasing $T_f$, they necessarily appear correlated in such a representation, but this of course does not imply any causal relationship~\cite{khomenko_relationship_2021}. In particular, our data seems to preclude a direct proportionality between the two quantities of the kind predicted in Ref.~\cite{ji2021toward}. This is especially true for the more realistic metallic glass model. Interestingly, while the density of TLS of the two models is very similar for glasses of equivalent stability, we find instead an increased density of QLM in the TLJ model. These observations suggest that generic QLM are poor predictors of TLS even though both families of localized excitations are similarly sensitive to the stability of the glass.  

\section{Conclusions}

\label{sec:conclusions}

Our analysis confirms that TLS are depleted with increasing glass stability in a second, more realistic and physically distinct glass-forming model~\cite{khomenko_depletion_2020}. The number of two-level systems  converges relatively quickly at $\sim0.3\%$ of the number of the double-well potentials identified during the classical exploration of the potential energy landscape. As the glass preparation temperature $T_f$ decreases, we find a slowly increasing ratio of the number of double wells over that of minima, which reflects a locally more connected energy landscape. The significant depletion of two-level systems with glass stability is in the end driven by the drastic reduction in the number of minima in a typical glass metabasin as $T_f$ decreases. The number of minima as a function of exploration time increases slowly as a sublinear power-law for all temperatures considered, with an exponent that decreases with decreasing $T_f$. This means that while the depletion of two-level systems remains robust, an exhaustive counting of both minima and defects remains elusive, even when exploration is strongly confined to a well-defined metabasin. 

The exploration of low energy glass metabasins reveals a significant degree of heterogeneity. In particular, we observe very large fluctuations, of several orders of magnitude, in the number of minima present in independent glass basins, even as $T_f \approx T_g$. An analysis of the metabasins from the perspective of connected pairs of IS shows a hierarchical structure of the energy landscape, as previously observed in Refs.~\cite{scalliet_nature_2019,liao2019hierarchical,artiaco2020exploratory}. The observation of these extremely strong sample-to-sample fluctuations in glass samples deserves further investigation~\cite{folena2022marginal,PhysRevLett.127.088002}. 

Comparing the continuously polydisperse soft-sphere system to a more realistic ternary model like TLJ, some important differences emerge as well. The absolute number of two-level systems, as estimated from the $n_0$ values for the TLJ model, are a bit closer to those obtained in real experiments, but remain larger by roughly an order of magnitude. While in the polydisperse model every particle is different, the ternary model contains particles of different types that are indistinguishable, which should decrease the possible number of defects, hence that of minima. From this point of view, glassy materials of increased chemical complexity should naturally exhibit more defects, while monodisperse or elemental counterparts of similar kinetic stability are likely to have fewer low-energy excitations. We observe only a modest effect though, which seems consistent with the experimentally observed quasi-universality of TLS density.

Another factor for the reduced density of two-level systems is the presence of attractive interactions and their influence on barrier heights. Control over the strength of attractive interactions may help modulate the properties of glassy materials through the reduction of low-energy excitations, but trade-offs might come in to play with other types of defects such as quasi-localized modes. Other interaction potentials could be studied as well, in order to better understand the role played by many-body interactions, anisotropy and dimensionality. 

When attempting to perform a quantitative comparison of the number of TLS observed in experiments and in computer simulations, one should keep in mind that in experiments landscape exploration is driven by quantum tunneling at $\sim 1$ K, which leads to a characteristic $\log(t)$ dependence of the TLS density, while in our computer study the landscape exploration is driven by classical thermal fluctuations at temperatures much higher than \SI{1}{\kelvin}, leading to a power-law growth of the number of minima with time. A direct comparison is then probably hindered by this important difference.

We have also explored the nature of the relationship between quasi-localized modes and two-level systems. We find the density of two-level systems to be correlated with that of quasi-localized modes, but we do not find them to be proportional or related by any causal link. At very low frequency, modes may no longer obey an $\omega^4$ scaling, although large uncertainties are present in our data. Interestingly, while the two models have a similar density of TLS, the density of QLM seems to increase for the TLJ model. This observation suggests that there is a diversity of defects present in glasses at low temperatures, which still evade an exhaustive classification. 

As perspectives for further numerical studies we would like to delineate three important research directions. First, to make our main conclusion linking glass stability to the depletion of the density $n_0$ of TLS even stronger, some of the approximations used to estimate $n_0$ should be removed from the simulations. The most ambitious task would be to explore the landscape and obtain the quantum splitting for candidate TLS, from a fully quantum-mechanical calculation using path-integral methods~\cite{marchi_pathintegral_1991, vaillant_tunneling_2018}. A second task would be to improve the computational tools used in this work to significantly increase the library of TLS through use of techniques in enhanced sampling and machine learning. Third, the discovery that glass metabasins contain such a large number of minima organised in a way that depends on glass stability, although hinted by previous work, requires a dedicated study to better quantify this evolution and understand how it impacts the physical properties of glassy and viscous liquid states. 

\begin{acknowledgements}
We thank E. Flenner, G. Folena, M. Ozawa, J. Sethna, S. Elliott, G. Ruocco and W. Schirmacher for useful discussions. This work was granted access to the HPC resources of MesoPSL financed by the Region Ile de France and the project Equip@Meso (reference ANR-10-EQPX-29-01) of the programme Investissements d’Avenir supervised by the Agence Nationale pour la Recherche. This project received funding from the European Research Council (ERC) under the European Union’s Horizon 2020 research and innovation program, Grant No. 723955 – GlassUniversality (FZ), and from the Simons Foundation (\#454933, LB, \#454955, FZ, \#454951 DR). CS acknowledges support from the Herchel Smith Fund and Sidney Sussex College, Cambridge.
\end{acknowledgements}

\section*{Data Availability Statement}

The data that support the findings of this study such as the statistics of sampled double-well potentials and equilibrium configurations of the ternary Lennard-Jones model are openly available at http://doi.org/10.5281/zenodo.7117711.

\appendix

\section{Large-scale behavior of $N_{DW}/N_{IS}$}

\label{appendix:nddwnis}

We discuss here the behavior of $N_{DW}/N_{IS}$ in the limit of a large number of samples and particles. Let us consider, as an example, a large system having $K$ independent localized excitations (i.e. DW potentials associated to the displacement of a few atoms). Note that within the TLS model, $K$ should be proportional to the number of atoms $N$ provided $N$ is large enough. Each DW can be represented by a one-dimensional local reaction coordinate $x_i$, with an associated potential taken, for the sake of illustration, 
to be of the form
\begin{equation}
    v_i(x_i)= \frac1{4!}x_i^4 + \frac12 \kappa_i x_i^2 - h_i x_i \ .
\end{equation}
If $\kappa_i^3 + 9 h_i^2/8<0$, this potential has two local minima, leading to a DW with energy splitting $\Delta V_i(\kappa_i,h_i)$.
Let us introduce a spin variable $\sigma_i=0, 1$ if $x_i$ is in the absolute, respectively local, minimum of $v_i(x_i)$. If the excitations are diluted and thus do not interact, the total energy of the system can be written as
\begin{equation}\label{eq:KHGPSnonint}
    V(\{x_i\}) = \sum_{i=1}^K v_i(x_i) \quad
    \Rightarrow \quad E_{IS}(\{\sigma_i\}) = \sum_{i=1}^K \Delta V_i \sigma_i \ .
\end{equation}
Such a system has $N_{IS}=2^K$ local minima, corresponding to all possible combinations of the $\sigma_i$, with energy $E_{IS}(\{\sigma_i\})$.
On the other hand, we have
$N_{DW}=K$, with each IS being connected to $K$ others: in fact, the MEP associated to a composite excitation of the form
$(\sigma_i=0, \sigma_j=0)\to(\sigma_i=1, \sigma_j=1)$ would be decomposed into elementary events, e.g. $(\sigma_i=0, \sigma_j=0)\to(\sigma_i=0, \sigma_j=1)\to(\sigma_i=1, \sigma_j=1)$, leading to the appearance of an intermediate local minimum in $(\sigma_i=0, \sigma_j=1)$ in the MEP.
This very simple argument thus suggests a relationship of the kind $N_{DW}\approx \log N_{IS}$.

In summary, for $K\propto N$ independent excitations in each glass, and for $N_g$ glasses, we would have in total $N_{IS}\sim N_g \exp(K)\sim N_g \exp(N)$, $N_{DW}\sim N_g K \sim N_g N$, and $N_{DW}/(N \, N_g)$ would then converge to a finite value, as in the TLS model.

Yet, elementary excitations deform the solid matrix, resulting in long-range elastic interactions, which would lead to a more accurate representation of the form (see e.g.~\cite{folena2022marginal} and references therein)
\begin{equation}
V(\{x_i\}) = \sum_{i=1}^K v_i(x_i) - \sum_{i<j} J_{ij} x_i x_j \ .
\end{equation}
For weak enough coupling $J_{ij}$, the system still has $N_{IS}=2^K$ local minima, but the DW profile of a single elementary excitation now depends on the state of all the other excitations in the system. As a result the MEP path connecting a given pair of IS is not necessarily decomposed into elementary transitions, and the number of DW can be as large as $N_{DW} \sim N_{IS}^2$ . Depending on the nature of elementary excitations and their interactions, one would then observe $\log N_{IS} \lesssim N_{DW} \lesssim N_{IS}^2$, and thus very different large scale behavior of the ratio $N_{DW}/N_{IS}$.
Extracting the relevant elementary excitations in this situation will obviously be cumbersome.

We can tentatively interpret the stability dependence of our numerical data as a smooth crossover between interacting DW for poorly stable glasses with a large concentrations of excitations to non-interacting ones in stable glasses where excitations become more dilute. 

\section{Data for polydisperse soft spheres}

\label{app:PSS}

\begin{figure}[!htb]
   	\includegraphics[width=\linewidth]{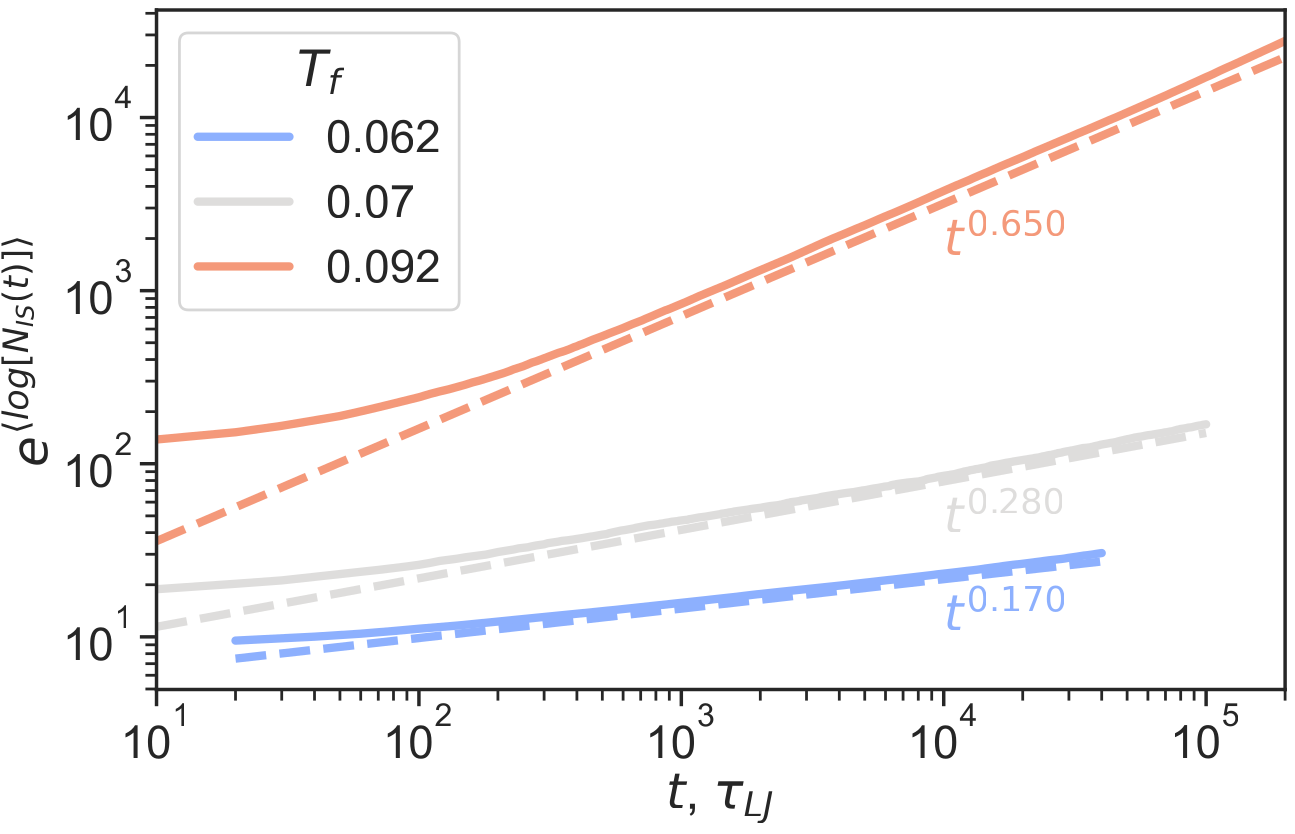}
   	\caption{\label{fig:NISPSS} Number of IS per glass averaged geometrically averaged over all glasses for polydisperse soft spheres. Dashed lines are guides for the eye, showing power-law scaling $\sim t^{\beta}$, with $\beta=0.17, 0.28, 0.65$ for $T_f=0.062, 0.07, 0.092$ respectively.}
      \end{figure}

In Fig.~\ref{fig:NISPSS} we show data for the scaling of inherent structures and DWs numbers with time, for the polydisperse soft-sphere glasses studied in~\cite{khomenko_depletion_2020}. We can see that the number of IS grows as a power-law, with scaling exponents changing from 0.65 to 0.17 when increasing the glass stability, which is qualitatively similar to the scaling of $N_{IS}(t)$ in TLJ glasses, shown in Fig.~\ref{fig:persist_minima}b. These exponents seem compatible with those obtained for the TLJ system.

\begin{figure}[!htb]
    \includegraphics[width=\linewidth]{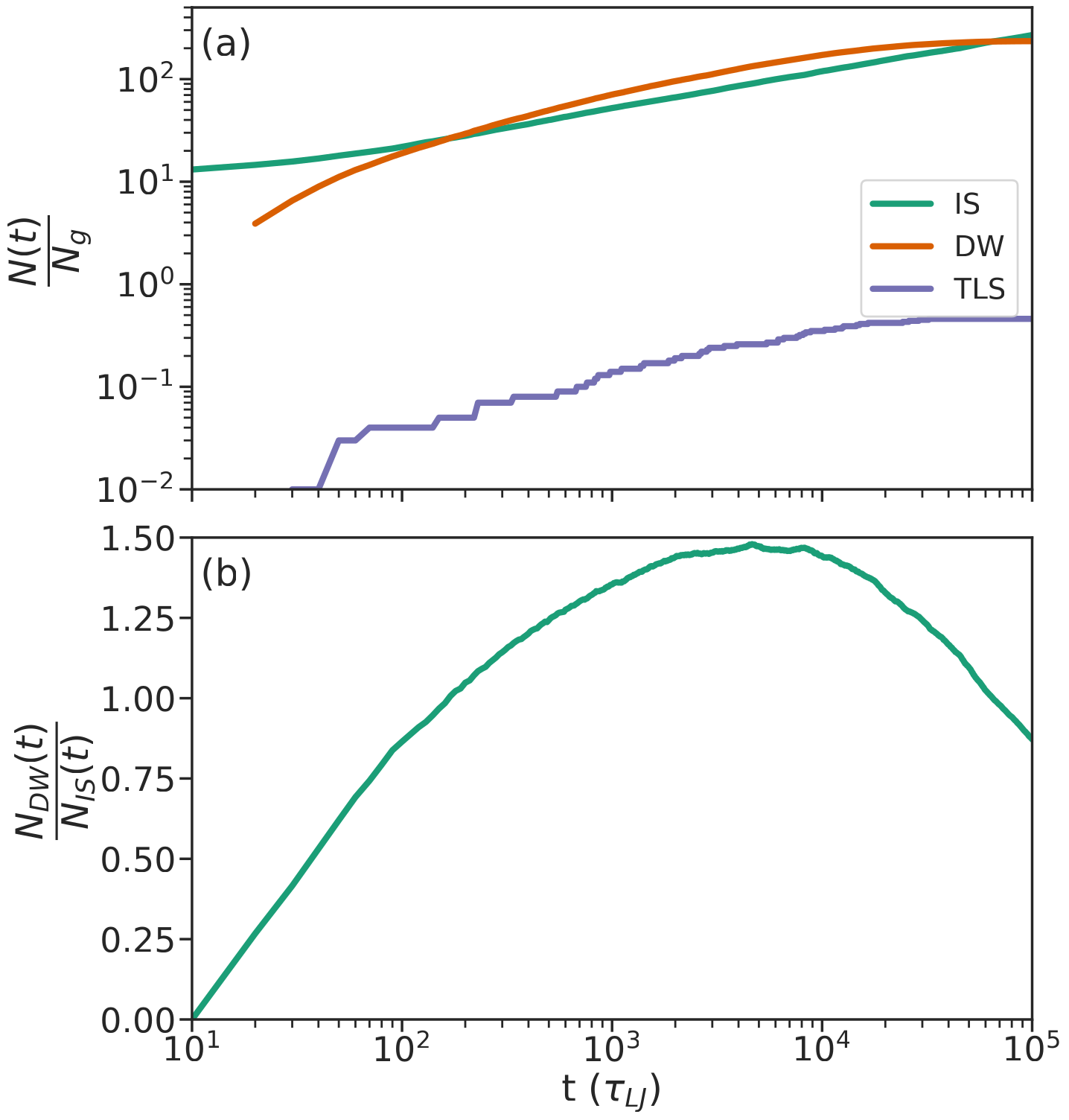}
    \caption{\label{fig:FtauPSS} Landscape exploration of polydisperse soft spheres prepared at $T_f=0.07$. (a) Number of minima (green), double-wells (orange) and TLS (purple) sampled as a function of time. (b) Ratio of number of DW to number of IS as a function of time. The two panels share the horizontal axis.}
\end{figure}

In Fig.~\ref{fig:FtauPSS} we show the number of detected DW as a function of time for PSS glasses. The number of DW grows slower than the number of IS, and scaling seems slower than a power-law. All these features are the same as those observed for TLJ glasses, as shown in Fig.~\ref{fig:trans_fpt}. A significant difference however is that at a similar glass stability and length of sampling there are significantly more DW per IS in the PSS model.

 \section{Relationship between the Hessian and the minimum-energy path}
\label{app:modesmep}

\begin{figure}[!htb]
    \centering
    \includegraphics[width=\linewidth]{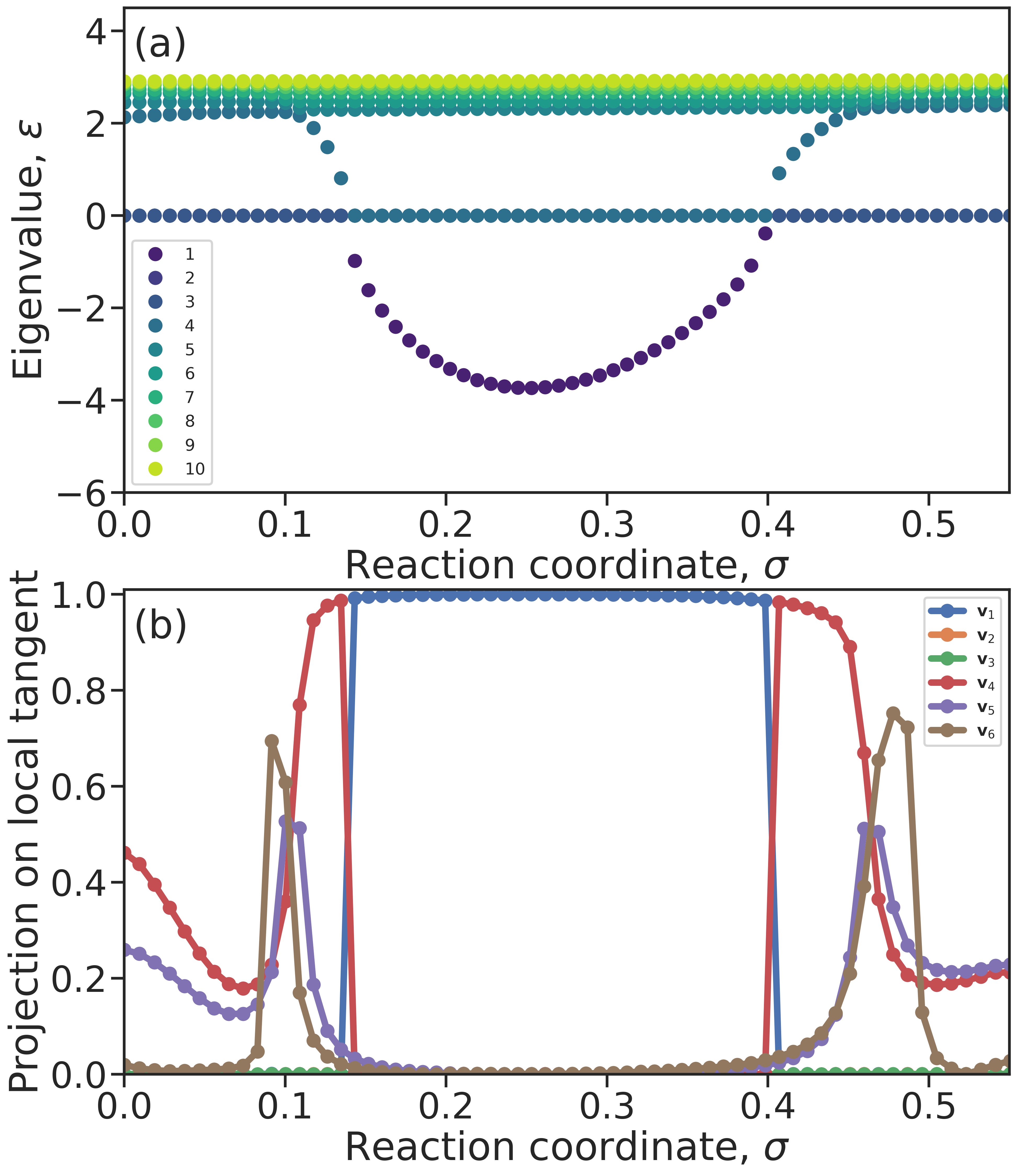}
    \caption{Eigenvalues and eigenvectors of the Hessian along the MEP of a typical TLS. (a) The first ten eigenvalues (in units of $\varepsilon$) along the MEP, the lowest eigenvalue is shown in \emph{purple} while the tenth mode is shown in \emph{yellow}. (b) Projection of the first six eigenvectors on the local tangent to the MEP. The first eigenvector is shown in \emph{blue}, while the next are shown in \emph{orange, green, red, purple} and \emph{brown}, respectively.} 
    \label{fig:eigMEP}
\end{figure}

We have studied the eigenvectors and eigenvalues of the Hessian for configurations along the MEP and these are shown in Fig.~\ref{fig:eigMEP}. We find a single eigenvalue (the lowest one) becomes negative along the MEP, indicating a first-order saddle point. Only several of the lowest frequencies change appreciably along the MEP. We also projected the first six eigenvector tangent to the MEP, and we find that the lowest mode contributes the most along the path, while contributions from several others become important only when approaching the minima. 

This validates the picture of a network of minima where each pair that corresponds to a TLS is connected by a 1D path, the MEP. Orthogonal to this path the dynamics is quasi-harmonic and is determined by a set of frequencies that are nearly independent from the reaction coordinate.

\section{Euclidean distance between minima and quantum splitting}
\label{app:dvsE}

\begin{figure}[!htb]
    \centering
    \includegraphics[width=\linewidth]{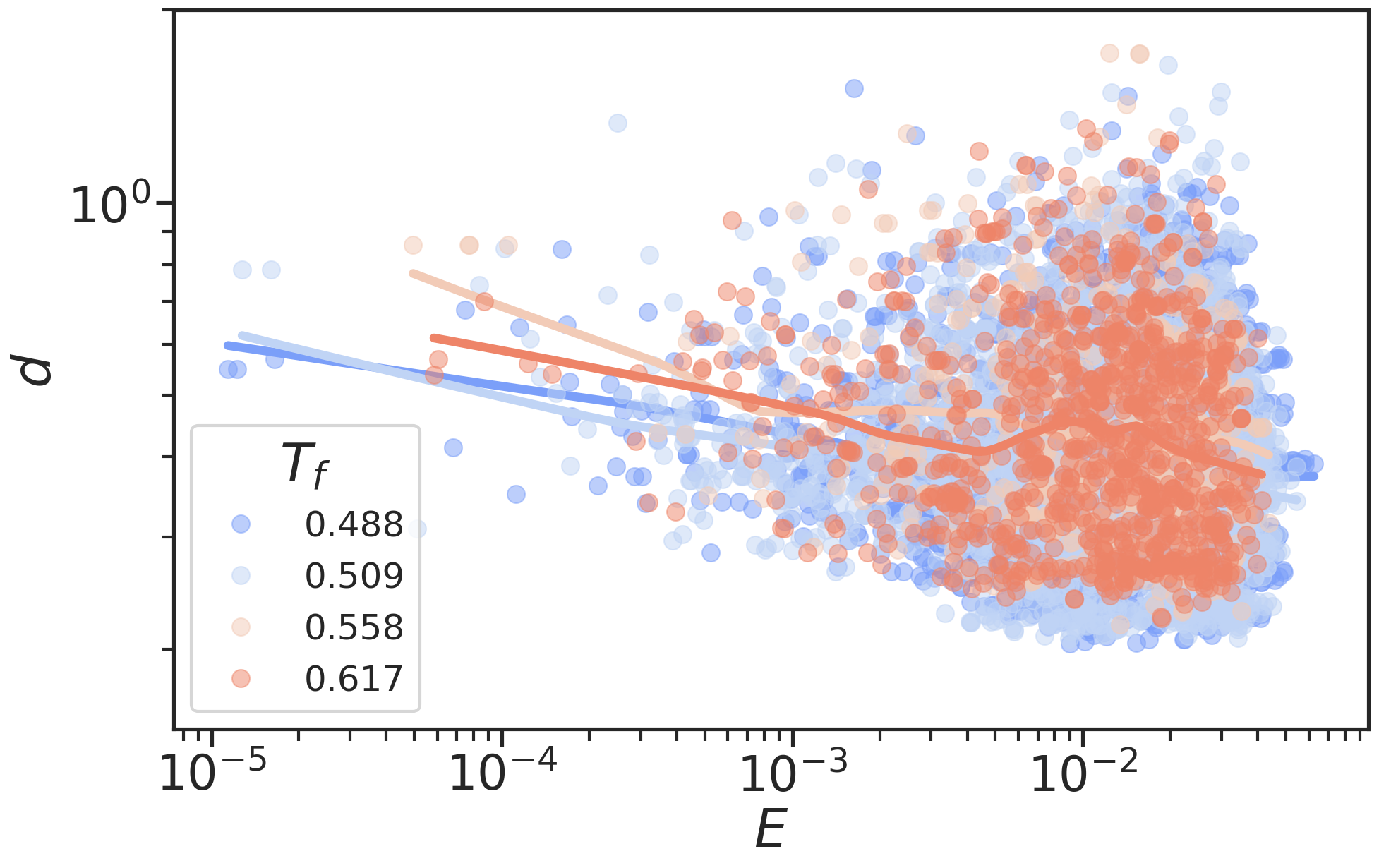}
    \caption{Euclidean distance $d$ versus splitting $E$. Colors code for different preparation temperatures $T_f$. Average values of $d$ at a given $E$ are shown as a solid line.}
    \label{fig:dvsE}
\end{figure}

The Euclidean distance between minima is a key quantity for tunneling DWs as it is strongly related to the tunneling rate. The scatter plot of Euclidean distance $d$ versus the quantum splitting $E$ is shown in Fig~\ref{fig:dvsE}. The Euclidean distance between IS tends to be larger at higher $T_f$.

\clearpage 

\bibliography{tls}

\end{document}